\documentclass[aps,nofootinbib,notitlepage,11pt]{revtex4-2}
\usepackage{amsmath,amssymb,slashed,color}
\usepackage{graphicx}
\usepackage{hyperref}
\usepackage{natbib}
\usepackage{bm}
\def\bal#1\gal{\begin{align}#1\end{align}}
\newcommand{\ball}[1]{\bal\label{#1}}
\newcommand{\eq}[1]{(\ref{#1})}
\newcommand{\fig}[1]{Fig.~\ref{#1}}
\renewcommand{\sec}[1]{Sec.~\ref{#1}}
\newcommand{\D}{{\rm d}}
\newcommand{\I}{{\rm i}}
\newcommand{\E}{{\rm e}}
\newcommand{\de}{\partial}
\newcommand{\h}[1]{\widehat{#1}}
\newcommand{\mean}[1]{\langle #1 \rangle}

\renewcommand{\vec}[1]{\ensuremath{\mathchoice				
		{\mbox{\boldmath$\displaystyle\mathbf{#1}$}}
		{\mbox{\boldmath$\textstyle\mathbf{#1}$}}
		{\mbox{\boldmath$\scriptstyle\mathbf{#1}$}}
		{\mbox{\boldmath$\scriptscriptstyle\mathbf{#1}$}}}}

\newcommand{\be}{\begin{equation}}
	\newcommand{\ee}{\end{equation}}         
\newcommand{\bea}{\begin{eqnarray}}
	\newcommand{\eea}{\end{eqnarray}}
\renewcommand{\b}[1]{{\bm #1}} 
\newcommand{\unit}[1]{\hat {{\bm #1}}} 
\newcommand{\aver}[1]{\left\langle #1 \right\rangle}
\DeclareMathOperator{\im}{\mathrm{Im}}

\begin{document}
	
\title{Photon radiation by relatively slowly rotating fermions in magnetic field}
	
\author{Matteo Buzzegoli}\email{mbuzz@iastate.edu}\affiliation{Department of Physics and Astronomy, Iowa State University, Ames, Iowa 50011, USA}

\author{Jonathan D. Kroth}\email{jdkroth@iastate.edu}\affiliation{Department of Physics and Astronomy, Iowa State University, Ames, Iowa 50011, USA}

\author{Kirill Tuchin}\email{tuchink@gmail.com}\affiliation{Department of Physics and Astronomy, Iowa State University, Ames, Iowa 50011, USA}
\author{Nandagopal Vijayakumar}\email{nvmg@iastate.edu}\affiliation{Department of Physics and Astronomy, Iowa State University, Ames, Iowa 50011, USA}

\begin{abstract}

We study the electromagnetic radiation by a fermion carrying an electric charge $q$ embedded in a medium rotating with constant angular velocity $\bm\Omega$ parallel or anti-parallel to an external constant magnetic field $\bm B$. We assume that the rotation is ``relatively slow"; namely, that the angular velocity $\Omega$ is much smaller than the inverse magnetic length $\sqrt{qB}$. In practice, such angular velocity can be extremely high. The fermion motion is a superposition of two circular motions: one due to its rigid rotation caused by forces exerted by the medium, another due to the external magnetic field.  We derive an exact analytical expression for the spectral rate and the total intensity of this type of synchrotron radiation. Our numerical calculations indicate very high sensitivity of the radiation to the angular velocity of rotation. We investigate its dependence on the sign of electric charge, direction of the magnetic field and the sense of rotation and show that the radiation intensity is  strongly enhanced if $q\bm B$ and $\bm \Omega$ point in the opposite direction and is suppressed otherwise.

\end{abstract}

\maketitle

\section{Introduction}
\label{sec:Intro}

The synchrotron radiation---electromagnetic radiation by charged fermions in the magnetic field---is a fundamental process of Quantum Electrodynamics that has many phenomenological applications in nearly every area of physics. In astrophysics it is emitted by relativistic particles in the extra-solar magnetic fields and it may be a source of jets generated by supermassive black holes, in relativistic nuclear physics it provides a mechanism of photon radiation from quark-gluon plasma and it serves as a diagnostic tool of the properties of condensed matter systems. 

According to the classical theory of synchrotron radiation, a charged particle of energy $E$ and charge $q$ in a magnetic field $B$ moves along a spiral trajectory with the synchrotron frequency $\omega_B= qBc/E$ and, by virtue of its acceleration, emits the electromagnetic radiation that was first computed by Schott in 1912\cite{schott1912electromagnetic}. The quantum theory of synchrotron radiation that takes into account the quantization of the fermion and photon fields was developed by Sokolov and Ternov \cite{book:SokolovAndTernov,book:Bordovitsyn} and has been extensively applied in astrophysics
\cite{Herold:1979,Herold:1982,Harding:1987,Baring:1988}. 

\begin{figure}[t]
    \centering
    \includegraphics[height=5.5cm]{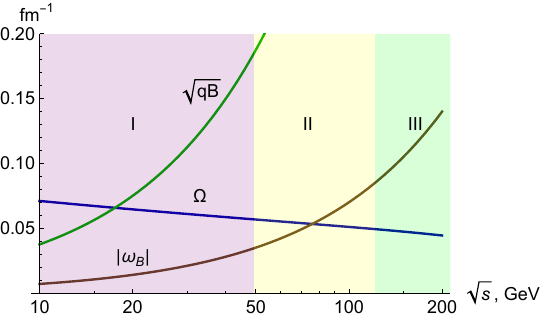}
    \caption{Typical values of the inverse magnetic length $\sqrt{qB}$, the absolute value of the synchrotron frequency $|\omega_B|$ and
    the vorticity $\Omega$ of the quark-gluon plasma versus the collision energy of the relativistic heavy-ions. Region I: fast rotation $\sqrt{qB}\sim \Omega$, the boundary conditions at $r=1/\Omega$ are very important. Region II: relatively slow rotation $\omega_B\sim \Omega\ll \sqrt{qB}$, the boundary conditions are not important, but the synchrotron radiation is affected by rotation. Region III: slow rotation $\Omega\ll \omega_B\ll \sqrt{qB}$, rotation can be neglected. An earlier version of this figure appeared in \cite{Tuchin:2021lxl}.}
    \label{fig:qgp}
\end{figure}

Often, the systems of charged fermions also rotate as a whole in an external magnetic field. In some exotic systems the angular velocity of rotation $\Omega$ is comparable or even exceeds the synchrotron frequency. In such systems the effect of rotation on the synchrotron radiation is significant. Table~\ref{table} provides several examples.  The most noteworthy example is the quark-gluon plasma produced in relativistic heavy-ion collisions. It has recently been observed that it is not only a subject of intense magnetic fields generated by the valence charges \cite{Kharzeev:2007jp,Skokov:2009qp,Voronyuk:2011jd,Ou:2011fm,Deng:2012pc,Tuchin:2013apa}, but also possesses 
huge vorticity, whose magnitude is comparable to the synchrotron frequency and which points in the same direction as the magnetic field \cite{Csernai:2013bqa,Csernai:2014ywa,Becattini:2015ska,Deng:2016gyh,Jiang:2016woz,Kolomeitsev:2018svb,Deng:2020ygd,Xia:2018tes}. This is illustrated in Fig.~\ref{fig:qgp}. 

\begin{table}\label{table}
\begin{tabular}{|c|c|c|c|}\hline
System & Magnetic field $B$ & Synchrotron frequency $\omega_B=\frac{eBc}{E}$ & Angular velocity $\Omega$ \\ \hline
Earth & $10^{-4}\,(\mathrm{Gs})/ 10^{-5}\,(\mathrm{eV^2})$ & $10^{-6}/E_\mathrm{eV}\,(\mathrm{eV})$ & $10^{-4}\mathrm{s}^{-1}/10^{-19}(\mathrm{eV})$\\
Dental drill & ---  & --- & $10^{4}\mathrm{s}^{-1}/10^{-11}(\mathrm{eV})$ \\
MRI & $10^{5}\,(\mathrm{Gs})/ 10^{4}\,(\mathrm{eV^2})$ & $10^{3}/E_\mathrm{eV}\,(\mathrm{eV})$ & --- \\
Reimann's nanoparticle \cite{PhysRevLett.121.033602}& ---& --- & $10^9\mathrm{s}^{-1}/10^{-6}(\mathrm{eV})$\\
Neutron star & $10^8-10^{15}\,(\mathrm{Gs})/$ & $10^6-10^{13}/E_\mathrm{eV}\,(\mathrm{eV})$ & $<10^3\,\mathrm{s}^{-1}/10^{-12}\,(\mathrm{eV})$\\
 &  $10^7-10^{14}\,(\mathrm{eV}^2)$ &  & \\
Quark-gluon plasma & $10^{14}-10^{18}\,(\mathrm{Gs})/$ & $10^{12}-10^{16}/E_\mathrm{eV}\,(\mathrm{eV})$ & $10^{24}\,\mathrm{s}^{-1}/10^{9}\,(\mathrm{eV})$\\
 & $10^{13}-10^{17}\,(\mathrm{eV}^2)$ &  & \\
\hline
\end{tabular}
\caption{The magnetic fields, the corresponding synchrotron frequencies and angular velocities for various systems. Notation $E_\mathrm{eV}$ means that energy $E$ is in units of eV. Note that the spectrum of the cosmic rays extends up to $E\sim 10^{21}$~eV. A more detailed look at the quark-gluon plasma is given in \fig{fig:qgp}. Useful unit conversions: $\mathrm{eV}^2=14.4\,\mathrm{Gs}$, $\mathrm{s}^{-1}=6.6\cdot 10^{-16}\,\mathrm{eV}$.  }
\end{table}

The rotating black hole is another natural candidate to observe the effect of rotation on synchrotron radiation. The angular velocities can be as high as $r_g^{-1}$, where $r_g$ is the gravitational radius. For a black hole of solar mass this amounts to $\Omega\sim 10^5\,\mathrm{s}^{-1}$. For comparison, the synchrotron frequency of a non-relativistic proton in a 1~Gs magnetic field is ten times smaller than $\omega_B(P)\sim 10^4\,\mathrm{s}^{-1}$. Fortunately, even less exotic systems may be sensitive to the rotation effects if the emitting particle energy is large enough, see Table~\ref{table}. A more mundane example is a modern dental drill that can rotate as fast as 13 kHz, which is roughly the same as $\omega_B(P)$. Observation of the effect of rotation on synchrotron radiation seems to be a realistic, albeit difficult, experimental problem.

These remarks motivate us to study the synchrotron radiation of rotating systems. Recently, we published a letter where we outlined our method and reported our first results \cite{Buzzegoli:2022dhw}. The goal of this article is to provide a through elaboration of our approach and present new results. We focus on the dynamics of the synchrotron radiation by a single rotating fermion, but we bear in mind future applications to quark-gluon plasma whose electromagnetic radiation may contain a significant synchrotron radiation component  \cite{Tuchin:2012mf,Yee:2013qma,Zakharov:2016kte,Wang:2020dsr,Wang:2020dsr}.  
Quantization of rotating quantum fields was discussed before in \cite{Letaw:1979wy,Bakke:2013sla,Konno:2012rt,Ayala:2021osy,Manning:2015sky}, and a variety of peculiar effects associated with rotation were considered in \cite{Anandan:1992zz,Post:1967oqg,Aharonov1973-AHAQAO,Anandan:1977ra,Staudenmann:1980uqe,Matsuo_2011,Fonseca:2017pnk,Chen:2021aiq,Chernodub:2017ref}. The effect of rotation and the magnetic field on bound states was also recently addressed by two of us \cite{Tuchin:2021lxl,Tuchin:2021yhy,Buzzegoli:2022omv}.

In relativistic heavy-ion collisions a charged particle moves together with the quark-gluon plasma. In the laboratory frame it rotates with angular velocity $\Omega$. We ignore the complexity of the electromagnetic field and instead focus on a qualitative analysis that assumes that in the plasma comoving frame there is only a uniform time-independent magnetic field $\b B$, while the electric field vanishes, as dictated by Ohm's law. We seek to compute the intensity of electromagnetic radiation emitted by such a particle as measured in the laboratory frame.

Unlike stationary systems, rotating systems must necessarily have finite radial size: causality demands that their radial extent has to be smaller than $1/\Omega$. Consider for example a wave function of a stationary system. It extends to spatial infinity even if the system is  well-localized, though it is strongly suppressed there. By applying the rotation operator $\exp(iJ_z\Omega)$, we cause the system to rotate. Now, however, the wave function is defined only at $r<1/\Omega$. A complete description of the system requires specifying the boundary conditions at $r=1/\Omega$. If the system size is comparable to $1/\Omega$, then the boundary conditions have a crucial effect on the system dynamics \footnote{Since any wave packet diffuses with time, the boundary conditions will eventually become essential for the system description.}. Conceptually, the role of the boundary is to balance the centrifugal and other fictitious forces that attempt to scatter the system to radial infinity. This is true even for classical systems as we discuss in \sec{sec:Classical}. For systems in thermal equilibrium, the boundary conditions are expected to have a significant effect on the equation of state. In this paper we assume that $1/\Omega$ is by far the largest radial distance,  which allows us to ignore the contributions at the boundary. This is the meaning of the term `relatively slow' rotation mentioned in the title.  However, the boundary effects are crucial for rapidly rotating systems. 
See \fig{fig:qgp} for an instructive example. The calculations in this paper are valid in regions II and III.

To obtain the wave functions of the rigidly rotating fermion system one can employ two  methods. One method consists in rotating the inertial frame solution of the Dirac equation to the rotating frame. This method was employed by Vilenkin to obtain the thermal Green's functions of the rotating system \cite{VilenkinQFT}. Another method consists in solving the Dirac equation in the rotating frame.
 This method has recently been employed in \cite{Chen:2015hfc,Mameda:2015ria} to study the combined effect of the magnetic field and rotation on the NJL model at finite temperature. The two methods are presumed to be equivalent at least as long as the system size is much smaller than the maximum radial distance $1/\Omega$ allowed by causality. The situation is more complex in the presence of a magnetic field.  The constant magnetic field in the stationary laboratory frame (as in a typical experiment with magnetic materials), transforms to a more complicated field configuration in the non-inertial rotating frame, see e.g.\ \cite{Tuchin:2021lxl}. However, if it is constant in the rotating frame (as in astrophysical applications and in heavy-ion collisions), then the situation is converse. We chose to consider the latter scenario for two reasons: (i) we are interested in eventually applying our results to relativistic heavy-ion phenomenology and  (ii) our results apply even in the former case in the non-relativistic limit, relevant in most condensed matter systems. Indeed, $B$ is invariant under rotation around its direction up to terms of order $1/c^2$.

We would like to stress that in our setup only the fermion field rotates, while the photon field does not.\footnote{One may envisage a different process, where the photon field rotates along with the fermion field in the magnetic field, e.g.\ the electromagnetic plasma in a tokamak. In this case one can employ the equivalence principle to first evaluate the intensity of the synchrotron  radiation in the instantaneously comoving inertial frame and then  transform it to the laboratory frame.} As a result, the fermion wave function depends on the angular velocity, while the photon wave function does not. In cylindrical coordinates this effect is accounted for by shifting the fermion energy by $-m\Omega$ where $m$ is an eigenvalue of the angular momentum operator $\h{J}_z$ \cite{deOliveira:1962apw,Hehl:1990nf}.

The boundary condition at $r=1/\Omega$ from the rotation axis necessarily induces quantization of the spectrum in the radial direction. Even though it is known how to take the exact boundary conditions into account \cite{Ambrus:2015lfr,Hortacsu:1980kv,Buzzegoli:2022omv}, they introduce cumbersome algebraic complications which are unnecessary in the regime studied here. As we have already indicated above, the calculation is greatly simplified in the slow rotation limit $\Omega\ll \sqrt{|qB|}$  since the characteristic extent of the fermion wave function in the radial direction is $1/\sqrt{|qB|}$. This approximation was also used in \cite{Chen:2015hfc}. This essentially eliminates the fermion current at the boundary and obviates the need for the boundary condition. We discuss this in more detail in \sec{sec:DiracEq}.

Our paper is organized as follows. In \sec{sec:Classical} we develop a classical approach to the synchrotron radiation of a rotating system. It illuminates some of the conceptual issues that otherwise may be misconstrued. In \sec{sec:DiracEq} we discuss the exact solution of the Dirac equation for rotating fermions in a constant magnetic field. The electromagnetic field is quantized in \sec{sec:Photons} in the basis of the Chandrasekhar-Kendall states. The differential and total radiation intensity is derived in \sec{sec:RadiationInt} and \sec{sec:total}. Our main analytical result is \eq{eq:DiffIntensity} for the differential radiation intensity.
The numerical procedure is described in \sec{sec:Numerics}.
The radiation spectrum is displayed in \fig{fig:spectrum}, the angular distribution of the intensity in \fig{fig:ThetaIntensityComparison}, and the total intensity in \fig{fig:total1}. We summarize in \sec{sec:summary}.

Throughout the paper $q$ denotes the  electric charge carried by the fermion, $B$ the magnetic field and $\Omega$ the angular velocity of rotation. 3-vectors are distinguished by bold face. We adopt the natural units $\hbar=c=1$ unless otherwise indicated.

\section{Warm-up: classical model}\label{sec:Classical}

Before we plunge into a fully quantum calculation, it is instructive to consider a classical model. This will provide us with a number of useful insights.

Consider a particle of negative charge $q=-|q|$ and mass $M=1$, such as an electron, embedded into a medium  rotating with constant angular velocity $\b\Omega = \Omega \unit z$. For example, this particle can be a quark in the quark-gluon plasma. Embedding means that the centrifugal force does not drive the particle to infinity in the rotation plane. If there were no magnetic field, the particle would be stationary in the frame rotating with the medium. However, due to the Lorentz force exerted on the particle by the constant magnetic field $\b B= B\unit z$ with $B>0$, it rotates counterclockwise in a circular orbit in the $(xy)$ plane with the angular velocity $\omega_B$. The particle trajectory in the rotating frame is 
\ball{Ac1}
&x_0=R-\varrho\sin(\omega_B t)\,,\quad 
y_0=\varrho\cos(\omega_B t)
\gal
where $(R,0)$ is the orbit center, $\varrho=V/\omega_B$ is the orbit radius, $V>0$ is the particle velocity and $\omega_B=|qB|\sqrt{1-V^2}$. Using the boost-invariance along the $z$-axis we chose a reference frame with $z=0$. 

\begin{figure}[ht]
\begin{tabular}{cc}
      \includegraphics[height=5cm]{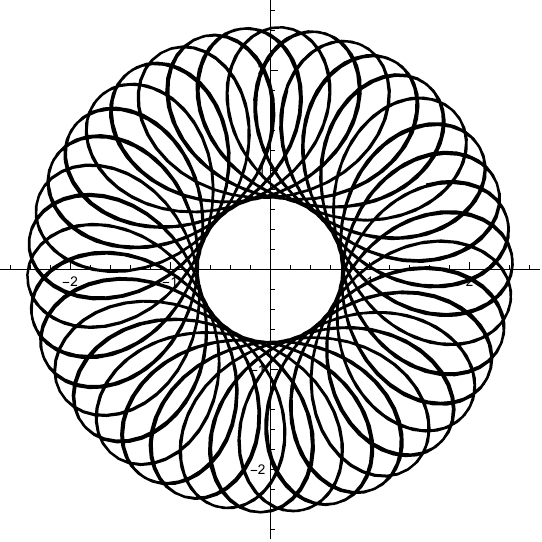} 
      &
      \includegraphics[height=5cm]{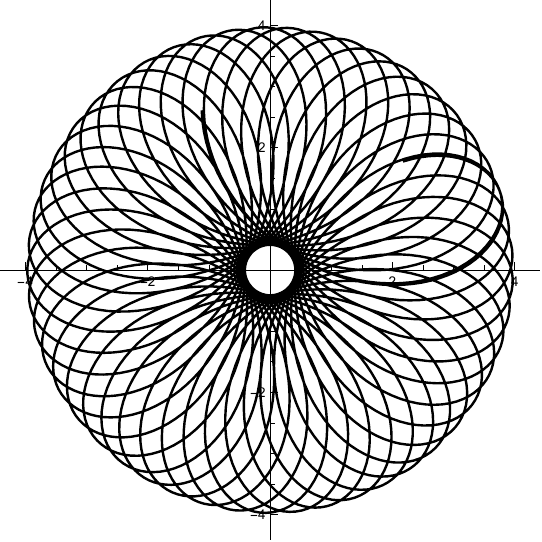} 
      \end{tabular}
  \caption{Trajectory of a classical particle of unit mass $M=1$, angular momentum $L=1$ and energy $E=1.5$ (left panel) or $E=2.5$ (right panel) in magnetic field $qB=-1$ embedded into rotating medium $\Omega=0.1$. The entire trajectory must be inside the rotating medium. $L$ and $E$ are defined in the following text.}
\label{figA2}
\end{figure}

Since the medium, together with the particle, rotates with angular velocity $\b \Omega$ with respect to the laboratory frame, the particle trajectory in the laboratory frame is determined by rotating $\eq{Ac1}$ through the angle $\Omega t$ about the $z$-axis. In the lab frame
\ball{Ac1.a}
x&=x_0\cos(\Omega t)-y_0\sin(\Omega t)\,,\quad
y=x_0\sin(\Omega t)+y_0\cos(\Omega t)
\gal
implying that 
\begin{subequations}\label{Ac3}
\bal
 &x=R\cos(\Omega t)-\varrho\sin((\omega_B+\Omega) t)\\
&y=R\sin(\Omega t)-\varrho\cos((\omega_B+\Omega) t)
\gal
\end{subequations}
In terms of the complex variable $\tilde x= x+iy$ \eq{Ac3} can be written compactly as
\ball{Ac5}
\tilde x= Re^{i\Omega t}+i\varrho e^{i(\Omega+\omega_B)t}
\gal
The particle trajectory \eq{Ac3} in the lab frame  is shown in \fig{figA2}. 

It can be shown that Eqs.~\eq{Ac3} satisfy the equation of motion:
\ball{Ac6}
\ddot {\b r}= -(2\Omega +\omega_B) \unit z\times \dot{\b r}+\Omega(\Omega+\omega_B)\b r\,.
\gal
The two terms on the right-hand-side are forces exerted on the particle. The first does no work ($\b f\cdot d\b r=0$) and is responsible for the circular motion. The second is the drag that is responsible for keeping the particle within the torus (as discussed above, if not for the drag our particle would spin out to $r\to \infty$). 

Now we wish to consider motion with given total energy $E$ and angular momentum projection on the $z$-axis $L$. (In the quantum case this corresponds to the stationary states in cylindrical coordinates.) To this end we need to determine the conserved quantities $E$ and $L$ and express the parameters $\varrho$ and $R$ through them. First, dot \eq{Ac6} with $\dot{\b r}$: 
\ball{Ac7}
\frac{d}{dt}\left\{ \frac{1}{2}\dot{\b r}^2-\Omega(\Omega+\omega_B)\frac{1}{2}\b r^2\right\}=0
\gal
Thus,
\ball{Ac8}
\mathcal{E}= \frac{1}{2}\dot{\b r}^2-\Omega(\Omega-\omega)\frac{1}{2}\b r^2
\gal
is a conserved quantity. It is natural to identify it with the non-relativistic kinetic energy. The relativistic energy is then
\ball{Ac9}
E=\frac{1}{\sqrt{1-2\mathcal{E}}}\,.
\gal
Noting that 
\bal
\b r^2&=\tilde x\tilde x^*= R^2+\varrho^2+\frac{2\omega_B \varrho R}{\Omega}\sin(\omega_B t)\,\label{Ac9.1}\\
\dot{\b r}^2&= \dot{\tilde x}\dot{\tilde x}^*=R^2\Omega^2+\varrho^2(\Omega+\omega_B)^2-2R\Omega \varrho(\Omega+\omega_B)\sin(\omega_B t)\,\label{Ac9.2}
\gal
and using \eq{Ac5} one gets
\ball{Ac10}
\mathcal{E}= \frac{\omega_B\varrho^2(\omega_B+\Omega)}{2}-\frac{\omega_B\Omega R^2}{2}\,.
\gal

Next, cross \eq{Ac6} with $\dot{\b r}$:
\ball{Ac12}
\frac{d}{dt}\left\{ \b r\times \dot{\b r}+(2\Omega+\omega_B)\frac{1}{2}r^2\unit z\right\}=0\,.
\gal
Thus,
\ball{Ac13}
\ell= {\unit z}\cdot ( \b r\times \dot{\b r})+(2\Omega+\omega_B)\frac{1}{2}r^2
\gal
is conserved. Clearly it is the projection of the non-relativistic angular momentum on the $z$-axis. The relativistic generalization is
\ball{Ac14}
L= E\ell\,.
\gal
One can use ${\unit z}\cdot ( \b r\times \dot{\b r})=\im (\tilde x^*\dot{\tilde x})$ to compute
\ball{Ac15}
\ell=\frac{\omega_B R^2}{2}-\frac{\omega_B \varrho^2}{2}\,.
\gal
Following the outlined program we now express $V=\varrho \omega_B$ and $R$  in terms of $E$ and $L$:
\bal
V^2&=1-\frac{1}{E^2}+\frac{2\Omega L}{E}\,\label{Ac17}\\
R^2&= \frac{1}{\omega_B^2}\left(1-\frac{1}{E^2}\right)+\frac{2(\Omega+\omega_B)L}{E\omega_B^2}\,,\label{Ac178}
\gal
where
\bal
\omega_B= |qB|\sqrt{1-V^2}=
|qB|\sqrt{\frac{1}{E^2}-\frac{2\Omega L}{E}}\,.
\gal

\begin{figure}[ht]
      \includegraphics[height=5cm]{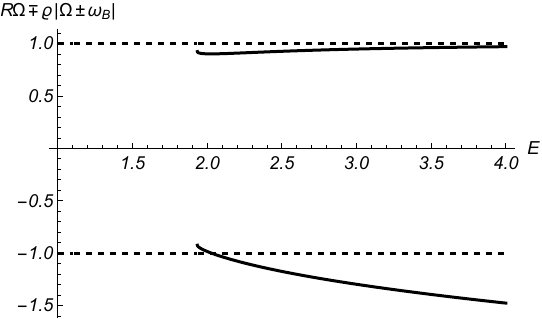} 
  \caption{Inequalities  \eq{Ac19} as a function of $E$; solid lines are the left-hand-sides. $qB=-1$, $L=-1$, $\Omega=0.1$, $M=1$. The first inequality of \eq{Ac19} is violated at about $E\approx 2$.}
\label{figA4}
\end{figure}

The parameters $E$ and $L$ are restricted by the causality constraint $\dot{\b r}^2\le 1$. In view of \eq{Ac9.2} it follows that 
$$
 \left(R\Omega\pm \varrho(\Omega+\omega_B)\right)^2\le 1
$$
and hence (assuming $R>0$ and $\varrho>0$)
\ball{Ac19} 
R\Omega-\varrho|\Omega+\omega_B|\ge -1\,,\quad R\Omega+\varrho|\Omega-\omega_B|\le 1 \,.
\gal
The left-hand and the right-hand  sides of \eq{Ac19} are plotted in \fig{figA4}.
Evidently, at any given $L$ the values of $E$ are restricted. Analogously, in the quantum case, the possible values of particle energy at any given value of the magnetic quantum number are restricted by causality.

The total intensity of the electromagnetic radiation emitted by a particle reads
\ball{Ac21}
W=\frac{2\alpha}{3}\frac{\ddot{\b r}^2-(\dot{\b r}\times \ddot{\b r})^2}{(1-\dot{\b r}^2)^3}
=\frac{2\alpha}{3}\frac{\ddot{\b r}^2(1-\dot{\b r}^2) +(\dot{\b r}\cdot \ddot{\b r})^2}{(1-\dot{\b r}^2)^3}\,,
\gal
where $\alpha=q^2/(4\pi)$. It is seen from \eq{Ac9.1} and \eq{Ac9.2} that $W$  is a periodic function of time with the period  $2\pi/\omega_B$. Therefore, its time average is
\ball{Ac22}
&\aver{W}=\frac{\omega_B}{2\pi} \int_0^{2\pi/\omega_B}Wdt\nonumber\\
& =\frac{\alpha}{3}
\frac{ 8 R^2 \varrho^2 \Omega ^3 (\omega_B +\Omega )^3-2 \left(R^2 \Omega ^2+\varrho^2 (\omega_B +\Omega )^2-1\right) \left(R^2 \Omega ^4+\varrho^2 (\omega_B +\Omega )^4\right)+R^2 V^2 \Omega ^2 (\omega_B +\Omega )^2}
{  \left(-2 \varrho^2 \left(R^2 \Omega ^2+1\right) (\omega_B +\Omega )^2+\left(R^2 \Omega ^2-1\right)^2+\varrho^4 (\omega_B +\Omega )^4\right)^{3/2}}.
\gal 

To study the limiting cases we fix the orbit center $R$ and its radius $\varrho=V/\omega_B$. Taking 
 $\Omega\to 0$ at fixed $\omega_B$ reduces \eq{Ac22} to
\ball{Ac23}
\aver{W}_{\Omega=0}= \frac{2\alpha}{3}(qB)^2\left(1-\frac{1}{E^2}\right)E^2\,,
\gal
which is a well-known result of the classical theory. Let us also record for the future reference the high energy limit of this formula in cgs units
\begin{align}\label{d35}
W_\mathrm{cl}=\frac{q^2}{4\pi}\frac{2 (qB)^2E^2}{3M^4}\,.
\end{align}
The opposite limit is $qB\to 0$ (hence $\omega_B\to 0$) at fixed $\Omega$. In this case in the  rotating frame the particle is at rest because there is no longer a Lorentz force exerted on it.  For simplicity consider $\varrho= 0$. Then \eq{Ac22} reduces to
\ball{Ac24}
\aver{W}_{qB=0}=\frac{\alpha}{3}\frac{R^2\Omega^4}{(1-R^2\Omega^2)^2}\,,
\gal
which is also a well-known result. 

\begin{figure}[ht]
      \includegraphics[height=5cm]{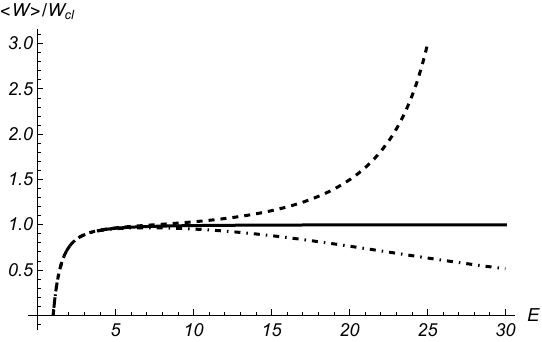} 
  \caption{$\aver{W}/W_\text{cl}$ as a function of $E$ at $qB=-1$, $L=1$, $M=1$, $\Omega=-10^{-5}$ (dash-dotted), $0$ (solid) and $10^{-5}$ (dashed). Recoil, neglected in this calculation, becomes important at $E\sim 4$.}
\label{figA3}
\end{figure}
The radiation intensity  is plotted in \fig{figA3} as a function of $E$. The first lesson is that the radiation intensity depends on the relative direction of the vectors $q\b B$ and $\b \Omega$: when they  point in the same direction the radiation is suppressed, whereas if they point in opposite directions, the radiation is enhanced. 

The divergence of $W$ in the case of anti-parallel $q\b B$ and $\b\Omega$ occurs when the denominator of \eq{Ac22} vanishes. This is, however, not a physical divergence, because it implies that the radiation energy exceeds the energy of the emitting particle, whereas Eq.~\eq{Ac22} was derived assuming that the radiated energy is negligible compared to $E$. 
The magnitude of the radiation reaction force is $W$ (along the direction of the particle velocity). It must be much smaller than the forces in the equation of motion \eq{Ac6}. 
We checked that for the parameters in \fig{figA3} the radiation reaction force reaches about a third of the Lorentz force when $E\sim 4$. Our classical model breaks down at this point. In the next section we begin developing a quantum model which naturally takes the recoil into account.

\section{Fermion wave function in magnetic field in rotating frame}\label{sec:DiracEq}

\subsection{Solution to the Dirac equation}
To represent a system rotating with angular velocity $\b\Omega$ in the $z$-direction, we consider a reference frame rotating with angular velocity $-\b\Omega$ about the $z$-axis and introduce a set of cylindrical coordinates $\{t,\,\phi,\,r,\,z\}$. 
The Dirac equation describing a fermion of mass $M$ and electric charge $q$ moving in the constant magnetic field 
 $\b B$ pointing in the $z$-direction reads
 \begin{align}\label{a1}
(i\gamma\cdot D-M)\psi  =0\,,
 \end{align}
 where
 \begin{align}\label{a2}
D_\mu=  \de_\mu + \Gamma_\mu + \I q A_\mu.
 \end{align}
Under the symmetric gauge $A^\mu=(0,-By/2,Bx/2,0)$, the only non-vanishing component of $\Gamma_\mu$,  related to the Christoffel symbols, is $\Gamma_0=-\Omega[\gamma^x,\gamma^y]/4$.  
 Eq.~(\ref{a1}) can be cast in the Schr\"odinger form 
 $i\partial_t\psi=\h{H}\psi$
with the Hamiltonian 
 \begin{align}\label{a4}
\h{H}=\gamma^0 \vec{\gamma}\cdot(\vec{p}-q\vec{A}) + \gamma^0 M  + \Omega \h{J}_z\,,
 \end{align}
where $\b p= -i\b\nabla$ and $\h{J}_z=-\I\de_\phi +\frac{\I}{2} \gamma^x\gamma^y$ are the operators of momentum and the total angular momentum correspondingly. 
The last term in (\ref{a4}) describes the effect of rotation.

The Hamiltonian $\hat{H}$ commutes with $\hat{p_z}$ and $\hat{J_z}$. Denote the corresponding eigenvalues as $E$, $p_z$ and $m$ respectively. Then
a solution of the Dirac equation in cylindrical coordinates has the following form, in the standard representation of the $\gamma$-matrices:
\begin{equation}
\label{eq:PsiSolutionGeneric0}
\psi(t,r,\phi,z)=\E^{-\I\epsilon E t}\frac{\E^{\I p_z z}}{\sqrt{L}}\frac{\E^{\I m \phi}}{\sqrt{2\pi}}  \left(\begin{array}{c}
	f_1(\rho) \E^{-\I\phi/2}\\
	f_2(\rho)\E^{+\I\phi/2} \\
	f_3(\rho)\E^{-\I\phi/2} \\
	f_4(\rho)\E^{+\I\phi/2}
\end{array}\right)\,,
\end{equation}
where $\epsilon=\pm 1$ labels the positive and negative-energy solutions and we introduced the dimensionless variable $\rho= \frac{|qB|}{2}r^2$. The radial functions $f_s(\rho)$, $s=1,\ldots,4$ must be determined by solving the Dirac equation.
Substituting (\ref{eq:PsiSolutionGeneric0}) into the Dirac equation and introducing the operators 
\begin{align}\label{eq:DefR1R2}
R_1 =& \sqrt{\frac{|qB|}{2}\rho}\left(2\frac{\D}{\D\rho}+\bar\sigma-\frac{m-\tfrac{1}{2}}{\rho} \right),\quad 
R_2 = \sqrt{\frac{|qB|}{2}\rho}\left(2\frac{\D}{\D\rho}-\bar\sigma+\frac{m+\tfrac{1}{2}}{\rho} \right),
\end{align}
where $\bar{\sigma}=\text{sgn}\left(qB\right)$, we find the equations satisfied by the radial functions 
\begin{align}
\label{eq:DiracRadial}
R_2 R_1 f_{1,3} =& - 2 |qB|\lambda f_{1,3}, \quad 
R_1 R_2 f_{2,4} = - 2 |qB|\lambda f_{2,4},
\end{align}
where 
\begin{equation}
\label{eq:lambda}
2 |qB|\lambda = \left(\epsilon E - \Omega m\right)^2 -M^2 -p_z^2
\end{equation}
defines the principal quantum number $\lambda$. 
 To solve Eqs.~(\ref{eq:DiracRadial}) we make a substitution 
\begin{equation}
f_{2,4} =\E^{-\rho/2}\rho^{\tfrac{m+1/2}{2}} u_{2,4}(\rho) \quad
f_{1,3} =\E^{-\rho/2}\rho^{\tfrac{m-1/2}{2}} u_{1,3}(\rho).
\end{equation}
 The auxiliary functions $u_s(\rho)$, $s=1,\ldots,4$ satisfy the Kummer equation
\begin{equation}
\label{eq:Kummer}
\rho\, u_s'' + (b_s-\rho) u_s' -a_s\, u_s=0
\end{equation}
with
\begin{align}
a_{2,4}=&\frac{1+\bar\sigma}{2} + \frac{1-\bar\sigma}{2}\left(m+\frac{1}{2}\right) - \lambda,\quad b_{2,4}=m+\frac{3}{2}, \\
a_{1,3}=&\frac{1-\bar\sigma}{2}\left(m+\frac{1}{2}\right) - \lambda,\quad b_{1,3}=m+\frac{1}{2}.
\end{align}

Solutions of (\ref{eq:Kummer}) that are finite at the origin $\rho=0$ can be expressed in terms of the confluent hypergeometric function 
\begin{equation}
u(\rho) = \mathcal{K} \frac{\,_1F_1 (a_s;\,b_s;\,\rho)}{\Gamma(b)},
\end{equation}
where $\mathcal{K}$ is a constant and $\Gamma$ is the gamma-function.
Apparently, the form of a solution depends on the sign $\bar\sigma$ of the product $qB$.

Another boundary condition must ensure that physical observables vanish at $r>1/\Omega$ where the proper time intervals in the rotating frame become imaginary. However, in the limit of slow rotation, when $\Omega\ll \sqrt{|qB|}$, the wave functions (\ref{eq:PsiSolutionGeneric0}) are exponentially small at $r=1/\Omega$. Thus, we can safely impose the boundary conditions at $\rho \to \infty$ instead. This effectively corresponds to unbounded motion. In both cases  $\bar\sigma=\pm 1$ we require that 
the functions $u_s$ vanish at  $\rho\to\infty$, which implies that $-a_s$ are non-negative integers \cite{book:SokolovAndTernov,NISTHandbook}. This implies that  the principal quantum number is also a non-negative integer $n$ \cite{book:SokolovAndTernov,NISTHandbook}:
\begin{equation}
\lambda = n = 0,1,2,\dots
\end{equation}
It follows from (\ref{eq:lambda}) that the energy dispersion relation reads
\begin{equation}\label{dr1}
(\epsilon E - \Omega m)^2 = 2|qB|n +p_z^2 +M^2.
\end{equation}
It is convenient to define the 
radial quantum number $a$ in place of $a_s$: 
\begin{align}
a_{1,3}= \frac{\bar\sigma+1}{2}\left(m-\frac{1}{2}\right)-a\,,\quad 
a_{2,4}= \frac{\bar\sigma+1}{2}\left(m+\frac{1}{2}\right)-a\,,
\end{align}
which implies
\begin{align}\label{def.a}
a=n+\bar\sigma m-\frac{1}{2}\,.
\end{align}
The radial wave functions read
\begin{align}
\label{eq:fRadial}
f_{1,3}^{\bar\sigma} = & \sqrt{|qB|}C^{\bar\sigma}_{1,3} I^{\bar\sigma}_1(\rho),\quad
f_{2,4}^{\bar\sigma} = \sqrt{|qB|}\I C^{\bar\sigma}_{2,4} I^{\bar\sigma}_2(\rho),
\end{align}
where 
\begin{align}\label{Lagguerre}
I^{\bar\sigma}_1(\rho) = I_{n-\frac{1-\bar\sigma}{2},a}(\rho),\quad
I^{\bar\sigma}_2(\rho) =  I_{n-\frac{1+\bar\sigma}{2},a}(\rho).
\end{align}
The Laguerre functions $I_{n,a}$ are related to the  generalized Laguerre polynomials $L_n^{(\alpha)}(z)$~\cite{NISTHandbook} as
\begin{equation}
\label{eq:def_I_and_rho}
    I_{n,a}(\rho)=\sqrt{\frac{a!}{n!}}\E^{-\rho/2} \rho^{\tfrac{n-a}{2}} L_a^{(n-a)}(\rho)\,.
\end{equation}

We normalize the eigenfunctions as
\begin{equation}\label{a10}
\int \psi^\dagger \psi\,\D^3 x = 1.
\end{equation}
In particular, the radial functions satisfy 
\begin{equation}
\sum_{s=1}^4\int_0^\infty  |f_s|^2 r dr=1,
\end{equation}
which implies the  constraint
\begin{equation}
\label{eq:Norm}
\sum_{s=1}^4 (C^{\bar\sigma}_s)^* C^{\bar\sigma}_s = 1.
\end{equation}

The coefficients $C_s^{\bar\sigma}$ can be computed by plugging (\ref{eq:PsiSolutionGeneric0}) back into the Dirac equation which produces 
\begin{align}
(\epsilon E - m \Omega \mp M) f_{1,3} +\I R_2 f_{4,2} - p_z f_{3,1} =& 0,\qquad 
(\epsilon E - m \Omega \mp M) f_{2,4} +\I R_1 f_{3,1} + p_z f_{4,2} = 0,
\end{align}
and using the identities
\begin{align}
R_1 I^{\bar\sigma}_1(\rho) =& + \bar\sigma \sqrt{2n |qB|} I^{\bar\sigma}_2(\rho)\,\qquad
R_2 I^{\bar\sigma}_2(\rho) = - \bar\sigma \sqrt{2n |qB|} I^{\bar\sigma}_1(\rho)
\end{align}
one obtains the algebraic equations
\begin{subequations}\label{eq:DiracEqC}
\begin{align}
(\epsilon E -\Omega\, m \mp M) C^{\bar\sigma}_{1,3} + \bar\sigma \sqrt{2n|qB|}C^{\bar\sigma}_{4,2} - p_z C^{\bar\sigma}_{3,1} =& 0 , \\
(\epsilon E -\Omega\, m \mp M) C^{\bar\sigma}_{2,4} + \bar\sigma \sqrt{2n|qB|}C^{\bar\sigma}_{3,1} + p_z C^{\bar\sigma}_{4,2} =& 0 .
\end{align}
\end{subequations}
%
Finally, the solutions of the Dirac equation take the form
\begin{equation}
\label{eq:PsiSolutionGeneric}
\psi_{\,n,\,a,\,p_z,\,\zeta}(x) =\E^{-\I\epsilon E t}\frac{\E^{\I p_z z}}{\sqrt{L}}\frac{\E^{\I m \phi}}{\sqrt{2\pi}} \sqrt{|qB|} \left(\begin{array}{c}
	C^{\bar\sigma}_1 I^{\bar\sigma}_1(\rho) \E^{-\I\phi/2}\\\I C^{\bar\sigma}_2 I^{\bar\sigma}_2(\rho)\E^{+\I\phi/2} \\
	C^{\bar\sigma}_3 I^{\bar\sigma}_1(\rho)\E^{-\I\phi/2} \\ \I C^{\bar\sigma}_4 I^{\bar\sigma}_2(\rho)\E^{\I\phi/2}
\end{array}\right),\quad n\geq 0,\; a \geq 0,
\end{equation}
where the coefficients $C_s^{\bar\sigma}$ satisfy (\ref{eq:DiracEqC}).
 Similar solutions in the chiral representation were previously discussed in \cite{PhysRevD.98.014017,PhysRevD.93.104052,Fang:2021mou}.

Eq.~(\ref{eq:Norm}) does not uniquely determine the coefficients $C_s$. Thus we are free to choose the functions (\ref{eq:PsiSolutionGeneric}) to be the polarization states of the fermion. There are two common choices of the polarization operators that commute with $\{\h{H},\, \h{J}_z,\, \h{P}_z \}$.

\subsection{Polarization states}

In Appendix~\ref{app:polarization}
we provide a detailed derivation of the helicity and spin magnetic moment operators that we use to characterize the polarization states. Here we give a brief summary of the results.

\subsubsection{Longitudinally polarized electrons - Helicity states}

We can choose the polarization states of the fermion to be the eigenstates of the helicity operator $\h{h}=\b \Sigma\cdot (\b p-q\b A)$.
To verify that it commutes with the Hamiltonian, it is useful to write it using (\ref{a4}) as $\h{h}=\gamma^5 (H - \Omega\h{J}_z) -\gamma^5 \gamma^0\, .$ 
The solutions of the Dirac equation that are also eignestates of the helicity operator obey the equation:
\begin{equation}\label{a20}
 \h{h}\, \psi = \zeta \sqrt{(E - \Omega m)^2-M^2}\; \psi,
\end{equation}
which results in
\begin{subequations}\label{eq:CoefficientsLongitudinalPol}
\begin{align}
C_1^{\bar{\sigma}} = & -\frac{\bar{\sigma}}{2} A_1 B_1,\,\quad 
C_2^{\bar{\sigma}} =   \frac{1}{2} A_1 B_2,\,\quad 
C_3^{\bar{\sigma}} =   -\frac{\bar{\sigma}}{2} A_2 B_1,\,\quad 
C_4^{\bar{\sigma}} =   \frac{1}{2} A_2 B_2,\\
A_1 = & \left(\frac{E-\Omega\, m + M}{E-\Omega\, m} \right)^{\frac{1}{2}},\quad
A_2 =   \zeta\left(\frac{E-\Omega\, m - M}{E-\Omega\, m} \right)^{\frac{1}{2}},\\
B_1 = & \left(1+\frac{\zeta\, p_z}{\sqrt{(E-\Omega\, m)^2-M^2}} \right)^{\frac{1}{2}},\quad
B_2 = \zeta\left(1-\frac{\zeta\, p_z}{\sqrt{(E-\Omega\, m)^2-M^2}} \right)^{\frac{1}{2}},
\end{align}
\end{subequations}
where $\zeta=\pm 1$.

Notice that for $M=0$, we have
\begin{equation}
\label{eq:ChiralStates}
\frac{C_3^{\bar{\sigma}}}{C_1^{\bar{\sigma}}}
=\frac{C_4^{\bar{\sigma}}}{C_2^{\bar{\sigma}}}=\frac{A_2}{A_1}=\zeta.
\end{equation}
Only for a massless field $\gamma^5$ commutes with the Hamiltonian and with $\widehat{J}_z$.
In that case requiring that $\psi$ is also an eigenstate of the chirality operator $\gamma^5$
leads to the conditions (\ref{eq:ChiralStates}). The helicity states above are equivalent to
the chiral states if the field is massless. For a massive field, the eigenstates of energy
do not have a definite chirality, even at the lowest Landau level $n=0$.

\subsubsection{Transversely polarized electrons - magnetic polarization states}

Another operator that commutes with the Hamiltonian and whose eigenstates we will use to characterize the polarization states is the spin magnetic moment  $\vec{\mu}$ defined as 
\cite{book:Bordovitsyn,book:SokolovAndTernov}:
\begin{align}\label{a28}
\vec{\mu}= \vec{\Sigma} - \frac{\I\gamma^0 \gamma^5}{2}\vec{\Sigma}\times (\vec{p}-q\vec{A})\,.
\end{align}
Its time evolution is governed by the equation $\dot{\b \mu}= \gamma^0\vec{\Sigma}\times q\vec{B}$. It follows that $\mu_z$ is a conserved quantity and its eigenvalues  
\begin{equation}\label{a30}
\h{\mu}_z \,\psi = \zeta\, \sqrt{(E - \Omega \, m)^2 -p_z^2}\; \psi,
\end{equation}
can be used to label the stationary states (\ref{eq:PsiSolutionGeneric}),
where $\zeta=\pm 1$. The corresponding eigenfunctions  are given by (\ref{eq:PsiSolutionGeneric}) with 
\begin{equation}
\label{eq:CoefficientsTransversePol}
\left(\begin{array}{c}
     C_1^{\bar\sigma} \\
     C_2^{\bar\sigma} \\
     C_3^{\bar\sigma} \\
     C_4^{\bar\sigma}
\end{array}\right) = \frac{1}{2\sqrt{2}}
\left(\begin{array}{c}
     -\bar{\sigma}B_3 \left( A_3 + A_4 \right) \\
     B_4 \left( A_4 - A_3 \right) \\
     -\bar{\sigma}B_3 \left( A_3 - A_4 \right)\\
     B_4 \left( A_4 + A_3 \right)\,
\end{array}\right),
\end{equation}
where
\begin{subequations}
\begin{align}
\label{eq:CoefficientsTransversePol2}
A_3 = & \left(\frac{E-\Omega\, m + p_z}{E-\Omega\, m} \right)^{\frac{1}{2}},\quad
A_4 =   \zeta\left(\frac{E-\Omega\, m - p_z}{E-\Omega\, m} \right)^{\frac{1}{2}},\\
B_3 = & \left(1+\frac{\zeta M}{\sqrt{(E-\Omega\, m)^2-p_z^2}} \right)^{\frac{1}{2}},\quad
B_4 = \zeta\left(1-\frac{\zeta M}{\sqrt{(E-\Omega\, m)^2-p_z^2}} \right)^{\frac{1}{2}}.
\end{align}
\end{subequations}
%

\subsection{Boundary conditions and causality}\label{sec:causality}

We have mentioned already that in the limit of slow rotation $\Omega\ll \sqrt{|qB|}$ the boundary condition can be set at the infinite radial distance from the rotation axis. This guarantees that the eigenfunctions are exponentially suppressed at $r>1/\Omega$. In particular, in this approximation, $\langle(r\Omega)^\kappa\rangle\ll 1$ for any positive $\kappa$ in a state with quantum numbers $n$ and $a$. For example, we can compute using (\ref{eq:PsiSolutionGeneric}) \cite{book:Bordovitsyn,book:SokolovAndTernov}:
\begin{equation}\label{a37}
\begin{split}
\mean{ r^2 } =& \int  \psi_{n,\,a,\,p_z,\,\zeta}^\dagger(x)\, r^2\, \psi_{n,\,a,\,p_z,\,\zeta}(x)\D^3 x\,  = \frac{2}{|qB|}\left[\left( n+a+\frac{1}{2}\right)-\mean{S_z}\right],
\end{split}
\end{equation}
where $\mean{S_z}$ is the average spin along the magnetic field.  For an unpolarized state we thus arrive at the condition $ 2\Omega^2\left( n+a+1/2\right)\ll |qB|$.
For other $\kappa$'s one obtains a different combination of $n$ and $a$, so, in general, causality demands that 
\begin{equation}\label{a39}
 n, a\ll N_{\rm caus}\equiv\frac{|qB|}{2\Omega^2} .
\end{equation}
This condition must be satisfied by the fermion state before and after the photon emission. 

We show in the next section, that the matrix elements for photon emission are proportional to the Laguerre functions $I_{n,n'}(x)$. Considering them as functions  of their order $n$, one can verify that the main contribution to the sum over $n'$ comes when $n\sim n'$. Thus, as long as $n$ of the initial state satisfies (\ref{a39}), the final state's $n'$ will satisfy it as well. The same is true of $a$ and $a'$. We discuss this in more detail in \sec{sec:Numerics} which deals with the numerical   analysis.

\section{Photon wave function}
\label{sec:Photons}

 The photon wave function in the radiation gauge $A^0=0$, $\bm\nabla \cdot \bm A=0$ is a solution to the wave equation 
\begin{align}\label{WaveEq}
(\b \nabla^2-\partial_t^2)\b A(x)=0\,.
\end{align}
Its solutions with given energy $\omega$ obey the vector Helmholtz equation
\begin{align}\label{HelmV}
(\b \nabla^2+\omega^2)\b A(x)=0\,.
\end{align}
The general solution to (\ref{HelmV}) is a linear combination of scaloidal, toroidal and poloidal fields that are obtained from the solution to the scalar Helmholtz equation 
\begin{align}\label{HelmS}
(\b \nabla^2+\omega^2)u(x)=0\,
\end{align}
as follows 
\begin{align}\label{STP}
    \vec{S} = \vec{\nabla}u\,, \qquad \vec{T} = \vec{\nabla}u \times \vec{a}\,, \qquad 
    \vec{P} = \frac{1}{\omega}\vec{\nabla} \times \vec{T},
\end{align}
where $\b a$ is an arbitrary vector.
Since we chose a gauge in which $\b A$ is divergenceless, it is spanned only by the toroidal and poloidal fields. 

Let $\b k$ be the photon momentum. Decompose it into components along the rotation axis and perpendicular to it: $\b k = k_z \vec{\hat{z}}+ \b k_\bot$. It follows from (\ref{WaveEq}) that $k^2= k_z^2+k_\bot^2= \omega^2$. 
The eigenfunctions of (\ref{HelmS}) are given by 
\begin{align}\label{b3}
u(\b x)=J_{\ell}\left(k_\bot r\right) e^{i\left(k_{z} z+\ell \phi\right)}\,.
\end{align}
Setting $\b a = \vec{\hat{z}}/k$ as the most convenient choice,  it follows from (\ref{STP}) that \cite{chandrasekhar1956force,chandrasekhar1956axisymmetric,chandrasekhar1957force,woltjer1958theorem}
\begin{align}
\vec{T}_{l,k_\perp,k_z}(\vec{x})&=\left\{\frac{\I l}{k r} J{_l}\left(k_{\perp} r\right) \vec{\hat{r}}-\frac{k_{\perp}}{k} J_{l}^{\prime}\left(k_{\perp} r\right) \vec{\hat{\phi}}\right\} \E^{\I\left(k_{z} z +l \phi\right)}\,\label{eq:Toroidal}\\
\vec{P}_{l,k_\perp,k_z}(\vec{x})&=\left\{\frac{\I k_z k_\perp}{k^2} J{_l}^{\prime}\left(k_{\perp} r\right) \vec{\hat{r}}-\frac{l k_{z}}{k^2 r} J_{l}\left(k_{\perp} r\right) \vec{\hat{\phi}} + \frac{k_\perp^2}{k^2} J_{l}\left(k_{\perp} r\right) \vec{\hat{z}}\right\} \E^{\I\left(k_{z} z +l \phi\right)}\,.
\label{eq:Poloidal}
\end{align}

A particular combination of $\b T$ and $\b P$ describes circularly polarized photons:
\begin{equation}
\label{eq:TwistedPhotonWF}
\vec\Phi_{h,l,k_\perp,k_z}(\phi,r,z)\equiv \frac{k}{k_\perp} \frac{1}{\sqrt{2}}
    \left(h\,\vec{T}_{l,k_\perp,k_z}(\phi,r,z) + \vec{P}_{l,k_\perp,k_z}(\phi,r,z) \right)\,,
\end{equation}
where $h=\pm 1$ labels right or left-handed photon states and the integer $l$ is the eigenvalue of the total angular momentum along $\vec{\hat{z}}$. These states are closely related to the twisted, or vortex, photon states \cite{PhysRevA.45.8185}\footnote{See also a recent review  \cite{Ivanov:2022jzh} that emphasizes applications in particle physics.}. Using the definitions (\ref{STP}) it can be shown that \begin{align}\label{b10}
\b \nabla \times \vec{\Phi}_{h, l, k_\bot, k_z}(\b x) = h k \vec{\Phi}_{h, l, k_\bot, k_z}(\b x)\,.
\end{align}
We normalize these states as 
\begin{align}\label{NormPhi}
      \int \vec{\Phi}_{h, l, k_\bot, k_z}(\b x) \cdot \vec{\Phi}_{h', l', k_\bot', k_z'}(\b x)
      d^3x =(2\pi)^2 \delta_{l l'}\delta_{h h'}\frac{\delta\left(k_\bot-k_\bot'\right)}{k_\bot}\delta\left(k_{z}-k_z'\right) .
\end{align}

The photon wave function, normalized to one particle per unit volume, reads 
\begin{align}\label{PhotonWaveF}
\vec{A}_{h,l,k_\perp,k_z}(x) =\frac{1}{\sqrt{2\omega V}}
    \vec\Phi_{h,l,k_\perp,k_z}(\phi,r,z) \E^{-\I\omega t}\,.
\end{align}
The general solution to the wave equation in cylindrical coordinates is then 
\begin{align}\label{b15}
\vec{A}(\vec{r}, t)=\sum_{l, h} \int \frac{d^{3} k}{(2 \pi)^{3}} a_{h,l,k_\bot,k_z} \frac{1}{\sqrt{2\omega V}} \vec{\Phi}_{h,l, k_\perp, k_{z}}(\vec{r}) \E^{-i \omega t}+ c.c.\,,
\end{align}
where the coefficients $a_{h,l,k_\bot,k_z}$ become the annihilation operators upon quantization of the electromagnetic field.

\section{Differential Radiation Intensity for \texorpdfstring{$qB<0$}{negative qB}}
\label{sec:RadiationInt}

In this and the following sections we compute the intensity of the synchrotron radiation in the case of $\bar\sigma=\text{sgn}\left(qB\right) =-1$, which makes the notation more compact. We then generalize to the case of $\bar\sigma =+1$ in the subsequent section where we also address the symmetries under the flip of the magnetic field and the angular velocity direction.  

The photon emission amplitude by a fermion of charge $q$ transitioning between two Landau levels is given by the $\mathcal{S}$-matrix element
\begin{align}\label{c1}
\mathcal{S}=(2\pi)\delta(E'+\omega-E)\frac{(-iq)}{\sqrt{2\omega V}}\int \bar\psi_{n',\,a',\,p_z',\,\zeta'}(\b x) \bm \Phi^*_{h,l,k_\bot,k_z}(\b x)\cdot \bm \gamma \psi_{n,\,a,\,p_z,\,\zeta}(\b x)\, d^3x\,,
\end{align}
where primed quantities refer to the final Landau level.
 The corresponding photon emission rate reads
\begin{align}\label{c3}
\dot w = \frac{(2\pi)q^2}{2\omega V}\delta(E'+\omega-E)\sum_{l,h}
\left|
\mean{\,\vec{j}\cdot\vec{\Phi}\,}\delta_{m',m-l}\frac{ 2\pi}{L} \delta(p_z-p_z'-k_z)
\right|^2 \frac{dk_zL}{2\pi} \frac{dk_\bot k_\bot \pi R^2}{2\pi}\,,
\end{align}
where we introduced a shorthand notation 
\begin{align}
\label{eq:MatrixElementsTP}
 \int \bar\psi_{n',\,a',\,p_z',\,\zeta'}(\b x) \bm \Phi^*_{h,l,k_\bot,k_z}(\b x)\cdot \bm \gamma \psi_{n,\,a,\,p_z,\,\zeta}(\b x)\, d^3x \equiv \mean{\,\vec{j}\cdot\vec{\Phi}\,}\delta_{m',m-l}\frac{ 2\pi}{L} \delta(p_z-p_z'-k_z)\,.
\end{align}
$L$ is the longitudinal extent of the system, and $R$ its transverse radius so that the system volume is $V=\pi R^2L$. The delta-symbols on the right-hand-side of (\ref{eq:MatrixElementsTP}) indicate the anticipated conservation of the $z$-components of the momentum and the angular momentum.

The radiation intensity, viz.\ energy radiated per unit time is given by $W=\dot w\,\omega$. The total radiation intensity is obtained by summing over  $n'$, $a'$, $\zeta'$ and integrating over $dp_z'L/(2\pi)$:\footnote{The integration over the longitudinal momentum can be done using the usual rule $\delta(p_z\to 0)=L/2\pi$ to obtain 
\begin{align}
\int \left|\frac{ 2\pi}{L} \delta(p_z-p_z'-k_z)\right|^2\frac{dp_z' L}{2\pi}=1\,.
\end{align}}
\begin{equation}
\label{eq:RadIntensity}
W_{n,a,p_z,\zeta} = \frac{q^2}{4\pi}  \sum_{n',a',\zeta'}\sum_{l,h}\delta_{m',m-l}\int  \left|
\mean{\,\vec{j}\cdot\vec{\Phi}\,}
\right|^2 \delta(\omega-E+E')dk_z k_\bot dk_\bot\,. 
\end{equation}
In practice it may be useful to introduce the polar angle $\theta$ as $k_z=\omega \cos\theta$. Then $k_\bot=\omega \sin\theta$ and the integration measure over the photon momentum is $dk_z dk_\bot = \omega d\omega d\theta$.


Now we proceed to evaluate the matrix elements $\mean{\,\vec{j}\cdot\vec{\Phi}\,}$ defined in (\ref{eq:MatrixElementsTP}). We decompose the photon wave function as
\begin{align}\label{Phistar}
    \vec{\Phi}^*(\phi,r,z)=\frac{1}{\sqrt{2}}\vec{\varphi}^*(r)\E^{-\I k_z z}
    \E^{-\I l \phi},
\end{align}
where using the Eqs. (\ref{eq:Toroidal}), (\ref{eq:Poloidal})
and (\ref{eq:TwistedPhotonWF}) and the identity
\begin{equation}\label{c10}
J_l'(k_\perp r)=\frac{1}{2}\left(J_{l-1}(k_\perp r)
    -J_{l+1}(k_\perp r)\right),
\end{equation}
we have
\begin{subequations}\label{phivector}
\begin{align}
\vec{\varphi}^*=  \varphi^*_r \,\vec{\hat{r}} +
    \varphi^*_\phi \, \vec{\hat{\phi}} +
    \varphi^*_z\, \vec{\hat{z}},
\end{align}
with
\begin{align}
\varphi^*_r(r) = & \I \frac{k_z}{k}\frac{J_{l+1}(k_\perp r)
    -J_{l-1}(k_\perp r)}{2}
    -\frac{\I h l}{k_\perp r}J_l(k_\perp r),\\
\varphi^*_\phi(r) = & \frac{h}{2}J_{l+1}(k_\perp r)
    -\frac{h}{2}J_{l-1}(k_\perp r)
    -\frac{l k_z}{k k_\perp r} J_l(k_\perp r),\\
\varphi^*_z(r) = & \frac{k_\perp}{k} J_l(k_\perp r).
\end{align}
\end{subequations}
To obtain the Cartesian components of the fermion current we use the explicit form of the wave functions (\ref{eq:PsiSolutionGeneric}). Denote
\begin{align}\label{chi.funct}
 \chi_{n,a}=\sqrt{|qB|}
    \left(\begin{array}{c}
         C_1 I_{n-1,a}(\rho) \E^{-\I\frac{\phi}{2}} \\
         \I C_2 I_{n,a}(\rho) \E^{\I\frac{\phi}{2}} \\
         C_3 I_{n-1,a}(\rho) \E^{-\I\frac{\phi}{2}} \\
         \I C_4 I_{n,a}(\rho) \E^{\I\frac{\phi}{2}}
    \end{array}\right)\,,
\end{align}
which yields  
\begin{subequations}\label{MatrEl}
\begin{align}
\chi_{n',a'}^{T} \gamma^0 \gamma^x \chi_{n,a} =& \I |q B|\left[
    K_1\E^{\I\phi}I_{n,a}I_{n'-1,a'} - K_2\E^{-\I\phi}I_{n-1,a}I_{n',a'}\right]\nonumber\\
    \equiv & |q B| (\I F_1^+ \E^{\I\phi} - \I F_1^- \E^{-\I\phi}),\\
\chi^T_{n',a'} \gamma^0 \gamma^y \chi_{n,a} =& |q B|\left[
    K_1\E^{\I\phi}I_{n,a}I_{n'-1,a'} +K_2\E^{-\I\phi}I_{n-1,a}I_{n',a'}\right]\nonumber\\
    \equiv & |q B| ( F_1^+ \E^{\I\phi} + F_1^- \E^{-\I\phi}),\\
\chi^T_{n',a'} \gamma^0 \gamma^z \chi_{n,a} =& |q B|\left[
    K_4 I_{n-1,a}I_{n'-1,a'} -K_3 I_{n,a}I_{n',a'}\right]
    \equiv |q B| F_3,
\end{align}
\end{subequations}
where we used the definitions
\begin{equation}
\label{eq:ABCD}
\begin{split}
    K_1 =& C_1' C_4 + C_3' C_2\,,\quad
    K_2 = C_4'C_1 + C_2' C_3,\\
    K_3 =& C_4' C_2 + C_2' C_4 \,,\quad
    K_4 = C_1'C_3 + C_3' C_1\, .
\end{split}
\end{equation}
Casting the scalar product $\vec{\varphi}^*\cdot\vec{\gamma}$ in the form
\begin{equation*}
\vec{\varphi}^*\cdot\vec{\gamma} =
    \left(\cos\phi \varphi_r^*(r) - \sin\phi\varphi_\phi^*(r) \right)\gamma^x
    +\left(\sin\phi \varphi_r^*(r) + \cos\phi\varphi_\phi^*(r) \right)\gamma^y
    + \varphi_z^*(r) \gamma^z
\end{equation*}
and using (\ref{phivector}) and (\ref{MatrEl}) the matrix element (\ref{eq:MatrixElementsTP}) becomes:
\begin{align}\label{AverJPHi}
\mean{\,\vec{j}\cdot\vec{\Phi}\,}=  \frac{|qB|}{\sqrt{2}}
   \int_0^\infty dr r
    \left[\left(\I F_1^+(r) - \I F_1^-(r)\right)\varphi_r^*(r)
        +\left(F_1^+(r) + F_1^-(r)\right)\varphi_\phi^*(r)
        +F_3(r) \varphi_z^*(r)\right]\,.
\end{align}

The integral over the radial variable involving $\varphi_z^*$ in (\ref{AverJPHi}) can be performed using the known formula \cite{book:SokolovAndTernov,book:Bordovitsyn,Kolbig:1996}
\begin{equation}
\label{eq:RadialInt}
\int_0^\infty J_{m-m'}\left(2(x\rho)^{1/2} \right) I_{n',a'}(\rho) I_{n,a}(\rho)\D\rho =
    I_{n,n'}(x) I_{a,a'}(x),
\end{equation}
and introducing the dimensionless variable
\begin{equation}
    x=\frac{k_\bot^2}{2|qB|}.
\end{equation}
Substituting $r = \sqrt{2\rho/|q B|}$ and $k_\bot=\sqrt{2|qB|x}$ we obtain
\begin{equation*}
\begin{split}
|qB| \int r\D r F_3(r) \varphi_z^*(r) =& \frac{k_\perp}{k}
   \int_0^\infty J_{m-m'}\left(2(x\rho)^{1/2} \right) F_3(\rho)\D\rho \\
=& \frac{k_\perp}{k}\left[K_4 I_{n-1,n'-1}(x)
    - K_3 I_{n,n'}(x)\right]I_{a,a'}(x)\,.
\end{split}
\end{equation*}
The other two radial integrals in (\ref{AverJPHi}) can be expressed in terms of the following four integrals 
\begin{subequations}\label{FourIntegrals}
\begin{align}
\mean{\mathcal{R}_1} = & \int \D \rho J_{m-m'-1}(k_\perp r)I_{n,a}(\rho)I_{n'-1,a'}(\rho),\\
\mean{\mathcal{R}_2} = & \int \D \rho \frac{J_{m-m'}(k_\perp r)}{k_\perp r}I_{n,a}(\rho)I_{n'-1,a'}(\rho),\\
\mean{\mathcal{R}_3} = & \int \D \rho J_{m-m'-1}(k_\perp r) I_{n-1,a}(\rho)I_{n',a'}(\rho),\\
\mean{\mathcal{R}_4} = & \int \D \rho \frac{J_{m-m'}(k_\perp r)}{k_\perp r}I_{n-1,a}(\rho)I_{n',a'}(\rho).
\end{align}
\end{subequations}
Altogether we have
\begin{align}\label{AverJPHI2}
\mean{\,\vec{j}\cdot\vec{\Phi}\,}=&
\frac{1}{\sqrt{2}}\sin\theta\left[K_4 I_{n-1,n'-1}(x) - K_3 I_{n,n'}(x)\right]I_{a,a'}(x)\nonumber\\
& + \frac{1}{\sqrt{2}}K_1\left(h-\cos\theta\right)\left[
    \frac{1}{2}I_{n,n'-1}(x)I_{a,a'}(x) +(m-m')\mean{\mathcal{R}_2}
    -\frac{1}{2}\mean{\mathcal{R}_1}\right]\nonumber\\
& - \frac{1}{\sqrt{2}}K_2\left(h+\cos\theta\right)\left[
    \frac{1}{2}I_{n-1,n'}(x)I_{a,a'}(x)+ (m-m')\mean{\mathcal{R}_4}
    -\frac{1}{2}\mean{\mathcal{R}_3}\right]\,
\end{align}

Thanks to a recurrence relation of the Bessel functions
\begin{equation*}
    \frac{2\nu}{z} J_{\nu}(z) = J_{\nu+1}(z) + J_{\nu-1}(z)\,
\end{equation*}
the integrals (\ref{FourIntegrals}) are not independent:
\begin{subequations}\label{IntegralRelations}
\begin{align}
(m-m')\mean{\mathcal{R}_2}-\frac{1}{2}\mean{\mathcal{R}_1} = \frac{1}{2}I_{n,n'-1}(x) I_{a,a'}(x)\,,\\
(m-m')\mean{\mathcal{R}_4} -\frac{1}{2}\mean{\mathcal{R}_3}= \frac{1}{2}I_{n-1,n'}(x) I_{a,a'}(x).
\end{align}
\end{subequations}
Substituting (\ref{IntegralRelations}) into (\ref{AverJPHI2}) we obtain the final expression for the matrix element 
\begin{align}\label{AverPHI3}
\mean{\,\vec{j}\cdot\vec{\Phi}\,}=& \frac{1}{\sqrt{2}}I_{a,a'}(x)\Big\{
\sin\theta\left[K_4 I_{n-1,n'-1}(x) - K_3 I_{n,n'}(x)\right]\nonumber\\
& + K_1\left(h-\cos\theta\right)I_{n,n'-1}(x) - K_2\left(h+\cos\theta\right)I_{n-1,n'}(x)
\Big\}.
\end{align}

Plugging (\ref{AverPHI3}) into (\ref{eq:RadIntensity})   yields the expression for the differential radiation intensity for a photon with circular polarization $h$:
\begin{align}\label{eq:DiffIntensity}
   W&_{n,a,p_z,\zeta}^{h} = \frac{q^2}{4\pi}  \sum_{n',a',\zeta'}\int \omega^2\sin\theta d\omega\, d\theta\, \delta(\omega-E+E') \frac{1}{2} I_{a,a'}^2(x) \nonumber\\
&\times \Big|
\sin\theta\left[K_4 I_{n-1,n'-1}(x) - K_3 I_{n,n'}(x)\right]
 + K_1\left(h-\cos\theta\right)I_{n,n'-1}(x) - K_2\left(h+\cos\theta\right)I_{n-1,n'}(x)
\Big|^2\,. 
\end{align}
By fixing $a'$ and $n'$ (and, by virtue of (\ref{def.a}), $m'$) one can also discuss the differential intensity for a photon with a given orbital angular momentum $l=m-m'$.

\section{Total radiation intensity for \texorpdfstring{$qB<0$}{negative qB}}
\label{sec:total}

The Landau levels on the initial and the final fermion are:
\begin{equation}\label{d3}
E = \sqrt{M^2 + 2n|qB|+p_z^2} + \Omega m,\quad
E' = \sqrt{M^2 + 2n'|qB|+p_z^{'2}} + \Omega m'.
\end{equation}
The energy and the longitudinal momentum  conservation conditions are 
\begin{align}\label{d5}
 \omega = E- E',\qquad 
p'_z = p_z - \omega \cos\theta .
\end{align}
Solving the above system for $p'_z$ and $\omega$ we obtain
\begin{align}
p'_z =& p_z - \omega_0 \cos\theta, \label{d7}\\
\omega_0 =& \frac{E-m'\Omega-p_z\cos\theta}{\sin^2\theta}
    \left\{ 1 - \left[ 1 - \frac{\mathcal{B} \sin^2\theta}{\left(E-m'\Omega - p_z\cos\theta \right)^2} \right]^{1/2}  \right\} , \label{d8}
\end{align}
where we defined
\begin{align}\label{d11}
\mathcal{B} =& 2 (n-n') |qB| - \Omega^2 (m-m')^2 + 2 (E-m'\Omega) \Omega(m-m').
\end{align}
These are the resonant frequencies at which photons are radiated by a rotating fermion in a magnetic field. 

It is convenient to perform a boost along the rotation axis into the frame where $p_z=0$.
Then the energy and momentum conservation conditions simplify 
\begin{align}
p'_z =& - \omega_0 \cos\theta, \label{d14}\\
\omega_0 =& \frac{E-m'\Omega}{\sin^2\theta}
    \left\{ 1 - \left[ 1 - \frac{\mathcal{B}\sin^2\theta}{(E-m'\Omega)^2} \right]^{1/2} \right\} .\label{eq:PhotonEnergyRotation}
\end{align}
The  delta-function in (\ref{eq:RadIntensity}) can be written as 
\begin{align}\label{DeltaFunction}
\delta(\omega - E + E') = \frac{\delta(\omega - \omega_0)}{\frac{\de(\omega -E + E')}{\de\omega}}
=\frac{\delta(\omega - \omega_0)}{1+\frac{\de E'}{\de\omega}}
\end{align}
with $\omega_0$ the solution (\ref{eq:PhotonEnergyRotation}) and
\begin{align}\label{d18}
\frac{\de E'}{\de\omega} = \frac{\de}{\de\omega}\left(\sqrt{ M^2 + 2n'|qB|+\omega^2\cos^2\theta}
+m'\Omega\right) = \frac{\omega\cos^2\theta}{E'-m'\Omega} .
\end{align}
Using (\ref{DeltaFunction}) in (\ref{eq:DiffIntensity}) and averaging over the initial fermion polarizations $\zeta$ we obtain
\begin{align}\label{I1}
    \frac{1}{2}\sum_\zeta W_{n,a,p_z=0,\zeta}^{h} = &\frac{q^2}{4\pi} \sum_{n',a'}\int_0^\pi d\theta\frac{\omega_0^2\sin\theta}{1+\frac{\omega_0\cos^2\theta}{E'-m'\Omega}}\Gamma^h_{n,a}(n',a',\theta),
\end{align}

where
\begin{align}\label{d22}
    \Gamma^h_{n,a}\equiv \frac{1}{2}\sum_{\zeta,\zeta'}
    \left| \mean{\,\vec{j}\cdot\vec{\Phi}\,}\right|^2
= \frac{1}{2}\left(\Gamma^{(0)}_{n,a} + h \Gamma^{(1)}_{n,a}\right) .
\end{align}
Summation over $\zeta$ and $\zeta'$ yields 
\begin{subequations}\label{Ks}
\begin{align}
\overline{K_1^2} \equiv& \frac{1}{2}\sum_{\zeta,\zeta'} K_1^2
    =\overline{K_2^2} = \overline{K_3^2} = \overline{K_4^2} \nonumber\\
    =&\frac{(E-m\Omega)(E'-m'\Omega)-M^2}{4(E-m\Omega)(E'-m'\Omega)},\label{d25}\\
\overline{K_1 K_2} =& 
      \overline{K_3 K_4} 
    =\frac{\sqrt{2n|qB|}\sqrt{2n'|qB|}}{4(E-m\Omega)(E'-m'\Omega)},\label{d26}\\
\overline{K_1 K_4} =& 
     -\overline{K_2 K_3}
    =-\frac{\sqrt{2n|qB|}\omega\cos\theta}{4(E-m\Omega)(E'-m'\Omega)},\label{d27}\\
\overline{K_1 K_3} =& 
    \overline{K_2 K_4}= 0.  \label{d28}
\end{align}
\end{subequations}
Substitution into (\ref{I1}) yields
\begin{align}\label{d30}
\Gamma^{(0)}_{n,a}= & I_{a,a'}^2(x)\Big\{
2 \overline{K_1^2}\left[ I_{n,n'-1}^2(x) + I_{n-1,n'}^2(x)\right]\nonumber\\
    &+\overline{K_1^2}\sin^2\theta\left[I_{n,n'}^2(x)
    +I_{n-1,n'-1}^2(x) - I_{n,n'-1}^2(x)-I_{n-1,n'}^2(x)\right]\nonumber\\
&-2\overline{K_1 K_2}\sin^2\theta \left[ I_{n,n'}(x)I_{n-1,n'-1}(x) + I_{n-1,n'}(x) I_{n,n'-1}(x) \right]\nonumber\\
&-2\overline{K_1 K_4} \sin\theta \cos\theta \left[I_{n-1,n'-1}(x)
   I_{n,n'-1}(x)+I_{n-1,n'}(x) I_{n,n'}(x)\right]\Big\}\\
    \label{d31}
\Gamma^{(1)}_{n,a}= & I_{a,a'}^2(x)\Big\{
    2 \overline{K_1^2} \cos\theta \left(I_{n-1,n'}^2(x)-I_{n,n'-1}^2(x)\right)\nonumber\\
&+2\overline{K_1 K_4} \sin\theta  \left(I_{n-1,n'-1}(x) I_{n,n'-1}(x)
    -I_{n-1,n'}(x) I_{n,n'}(x)\right)\Big\}.
\end{align}
%

\section{Radiation intensity at \texorpdfstring{$qB>0$}{positive qB}}
\label{sec:rad.posit.qB}

Thus far in Secs.~\ref{sec:RadiationInt},\ref{sec:total} we considered $qB<0$. To obtain the radiation intensity for $\bar\sigma=\text{sgn}\left(qB\right)=+1$ we keep the same reference frame and make the following changes:
\begin{itemize}
	\item Because of the change of sign of $C^{\bar\sigma}_{1,3}$ in (\ref{eq:CoefficientsLongitudinalPol}) or (\ref{eq:CoefficientsTransversePol}) and (\ref{eq:CoefficientsTransversePol2}):
		\begin{equation}
            K_1 \to -K_1,\quad K_2 \to -K_2,
		\end{equation}
    \item The definition of $a$ differs for opposite $\bar\sigma$, see Eq. (\ref{def.a}).
	\item Because of the change of  $I_{n,n'}$ function indices, see \eq{Lagguerre}, 
		\begin{equation}
			n \to n-1,\quad n-1\to n,\quad\text{and the same for }n'.
		\end{equation}
\end{itemize}
Following the same steps described in Sec.~\ref{sec:RadiationInt}, we eventually obtain
\begin{align}\label{eq:DiffIntensityMinus}
   \frac{dW_{n,a,p_z,\zeta}^{h\,\bar\sigma=+}}{d\omega} =& \frac{q^2}{4\pi}  \sum_{n',a',\zeta'}\delta_{m,m'+l}
   \int_0^\pi \omega^2\sin\theta  d\theta\, \delta(\omega-E+E') I_{a,a'}^2(x) \nonumber\\
&\times \Big|
\sin\theta\left[K_4 I_{n,n'}(x) - K_3 I_{n-1,n'-1}(x)\right]
 + K_1\left(h-\cos\theta\right)I_{n-1,n'}(x) \nonumber\\
 &- K_2\left(h+\cos\theta\right)I_{n,n'-1}(x)\Big|^2\,.
\end{align}
In general, comparing Eq.~(\ref{eq:DiffIntensityMinus}) with Eq.~(\ref{eq:DiffIntensity}), we can write
\begin{equation}\label{eq:DiffIntensitysigma}
   \frac{dW_{n,a,p_z,\zeta}^{h\,\bar\sigma}}{d\omega} = \frac{q^2}{4\pi}  \sum_{n',a',\zeta'}\delta_{m,m'+l}
   \int_0^\pi \omega^2\sin\theta  d\theta\, \delta(\omega-E+E') I_{a,a'}^2(x) \mathcal{W}^{h,\bar\sigma}_{n,n'}(x,\theta),
\end{equation}
where
\begin{equation}
\begin{split}
\mathcal{W}^{h,\bar\sigma}_{n,n'}(x,\theta) = \Big|& \sin\theta\left[K_4 I^{\bar\sigma}_{-,-}(x) - K_3 I^{\bar\sigma}_{+,+}(x)\right] 
 + K_1\left(h-\cos\theta\right)I^{\bar\sigma}_{+,-}(x)\\
 &- K_2\left(h+\cos\theta\right)I^{\bar\sigma}_{-,+}(x)\Big|^2\,,
\end{split}
\end{equation}
and
\begin{equation*}
I^{\bar\sigma}_{\pm,\pm}(x) = I_{n-\tfrac{1\pm\bar\sigma}{2},n'-\tfrac{1\pm\bar\sigma}{2}}(x) .
\end{equation*}

The total radiation intensity is
\begin{align}\label{eq:TotalInt}
   W_\text{tot}\equiv \frac{1}{2}\sum_\zeta W_{n,a,p_z=0,\zeta}^{h,\bar\sigma} = &\frac{q^2}{4\pi} \sum_{n',a'}\int_0^\pi d\theta\frac{\omega_0^2\sin\theta}{1+\frac{\omega_0\cos^2\theta}{E'-m'\Omega}}\frac{1}{2}\left(\Gamma^{(0)}_{n,a} - h\bar\sigma \Gamma^{(1)}_{n,a}\right)
\end{align}
where $\Gamma^{(0)}_{n,a}$ and $\Gamma^{(1)}_{n,a}$ are the same as in Eqs. (\ref{d30}) and (\ref{d31}).
%
%

Let us now compare the synchrotron radiation of a particle of charge $q$ moving in the magnetic field $\b B$ in rotating frame $\b \Omega$ and the same particle moving in the magnetic field $-\b B$ in rotating frame $-\b \Omega$. Flipping both the direction of the magnetic field and the angular velocity of rotation changes the sense of the rotational motion about the symmetry axis. Thus the total radiation intensity should be the same. Indeed, Eq.~(\ref{eq:TotalInt}) implies that
\begin{equation}
\label{eq:chargeTransform}
\frac{1}{2}\sum_{\zeta,\zeta'}\mathcal{W}^{h,\bar\sigma}_{n,n'}(x,\theta)
    = \frac{1}{2}\sum_{\zeta,\zeta'}\mathcal{W}^{-h,-\bar\sigma}_{n,n'}(x,\theta).
\end{equation}
With this in mind we can ask if there is a difference between the radiation summed over all helicities and spins from a particle in two magnetic fields with opposite direction (or, equivalently, from two
particles of opposite charge in the same field). That is, we want to know the difference
\begin{equation}
\Delta_{\bar\sigma} 
= \sum_{\zeta,\zeta',h}\left(
	\frac{dW_{n,a,p_z,\zeta}^{h\,+}}{d\omega} - \frac{dW_{n,a,p_z,\zeta}^{h\,-}}{d\omega}\right).
\end{equation}
When $\Omega=0$, the energy $E$ does not depend on $m$ and we can use the identity $\sum_{a'}  I_{a,a'}^2(x)=1$ ~\cite{book:SokolovAndTernov} to show, using the previous transformation (\ref{eq:chargeTransform}), that $\Delta_{\bar\sigma}=0$.
%
%

In contrast, the energy of a rotating system depends on $m$. Since fixing $n$ and $a$ while flipping $\bar\sigma$ results in opposite $m$, see (\ref{def.a}), to have the same energy we also need to change the sign of $\b\Omega$.
Only by changing the sign of $\b\Omega$ can we obtain the same expressions and we have
\begin{equation}
\sum_{h,\zeta,\zeta'}\left(\frac{dW_{n,a,p_z,\zeta}^{h\,+}}{d\omega}(+\b\Omega) - \frac{dW_{n,a,p_z,\zeta}^{h\,-}}{d\omega}(-\b\Omega)\right)=0,
\end{equation}
as expected.

\section{Numerical results}\label{sec:Numerics}

In this section we present the procedure and the results of the numerical calculation of the synchrotron radiation intensity for $\bar \sigma=\text{sgn}\left(qB\right) =-1$, e.g.\ an electron $q=-1$ in the magnetic field pointing in the $z$-direction $B>0$.
The radiation intensity for a positive charge can be easily obtained using the results of the previous section, by simply inverting the sense of rotation.

\subsection{Frequency spectrum}

\begin{figure}[ht]
      \includegraphics[width=0.45\textwidth]{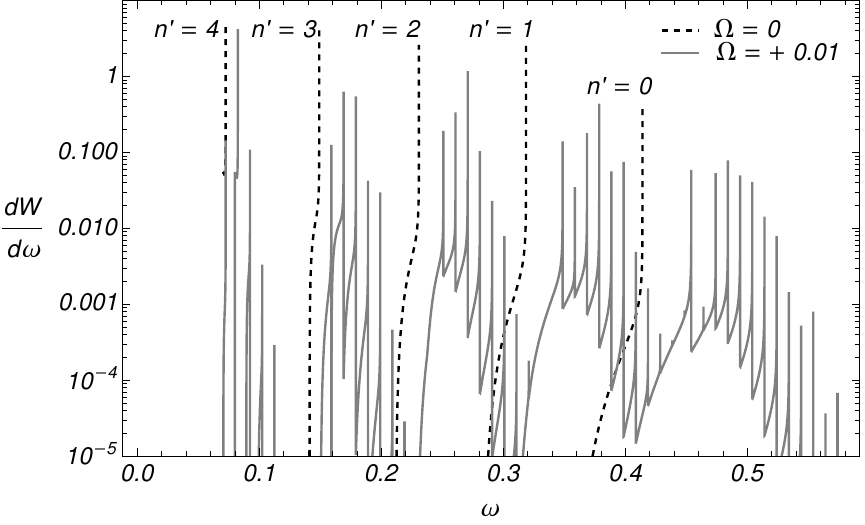}
      \includegraphics[width=0.45\textwidth]{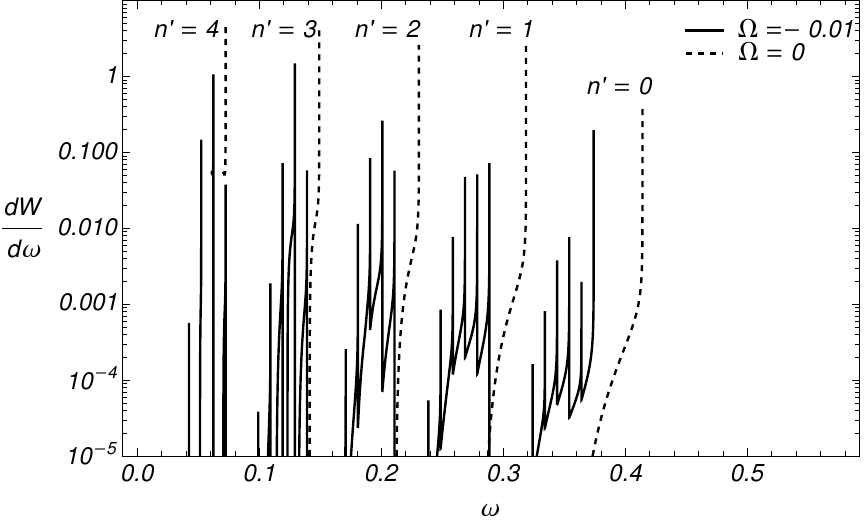}
  \caption{The spectrum of synchrotron radiation (\ref{eq:DiffIntensityInt}). Magnetic field strength: $qB=-0.1$. Initial quantum numbers: $n=5$, $a=1$, $m=7/2$ and $p_z=0$. The spectrum is summed over the final quantum numbers $0\le a'\le 50$ and $h=\pm 1$. Solid black lines: $\Omega=-0.01$ (corresponding to $E=1.379$), dashed lines: $\Omega=0$ (corresponding to $E=1.414$), and solid gray lines $\Omega=0.01$ (corresponding to $E=1.449$). Our units: $\hbar = c = M = 1$. }
\label{fig:spectrum}
\end{figure}

We derive the frequency spectrum by explicitly integrating over the angle $\theta$ in \eq{eq:DiffIntensity} using the  delta-function. The argument of the delta function has two roots $\theta_\pm$. The frequency spectrum is then the sum of the intensity expressions for the two angles:
\begin{align}\label{eq:DiffIntensityInt}
   \frac{dW_{n,a,p_z,\zeta}^{h}}{d\omega} = \frac{q^2}{4\pi}  \sum_{n',a',\zeta',\theta_\pm}&\frac{E'-m'\Omega}{\cos\theta_\pm}I_{a,a'}^2(x) 
 \Big\{
\sin\theta_\pm\left[K_4 I_{n-1,n'-1}(x) - K_3 I_{n,n'}(x)\right]\nonumber\\
 +& K_1\left(h-\cos\theta_\pm\right)I_{n,n'-1}(x) - K_2\left(h+\cos\theta_\pm\right)I_{n-1,n'}(x)
\Big\}^2\,,
\end{align}
where
\begin{align}\label{eq:theta0}
\cos\theta_\pm= \pm  \frac{1}{\omega}\sqrt{(E-\omega-m'\Omega)^2-2n'|qB|-M^2}\,.
\end{align}
\fig{fig:spectrum} exhibits a typical synchrotron radiation spectrum. For comparison we also plotted the spectrum emitted by the non-rotating fermion. While the spectrum of the non-rotating fermion depends only on the principal  quantum number $n'$, the spectrum of the rotating fermion is split in many lines having different $a'$, as expected from the energy shift caused by rotation. Moreover, the positions of the spectral lines are shifted toward larger (smaller) values of $\omega$ for positive (negative) sense of rotation and their heights are diminished in comparison with the non-rotating spectrum. This indicates that the radiation intensity is enhanced (suppressed) for $\bar\sigma =-1$  and positive (negative) sense of rotation.

\subsection{Total intensity}
 
\begin{figure}[t]
      \includegraphics[width=4in]{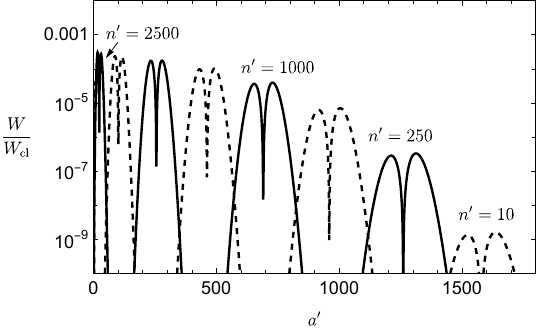}
  \caption{The intensity of photon emission normalized to the classical intensity as a function of $n'$ and $a'$ with $n = 3000, a = 1, qB = -10^{-2},$ and $\Omega = 10^{-5}$ (corresponding to $E = 7.840$). Each curve has a different $n'$. Alternating line styles emphasize a different value of $n'$. Our units: $\hbar = c = M = 1$.
  }
\label{fig:aprimeDependence}
\end{figure}

We calculate the total intensity numerically by summing over the quantum numbers $n'$ and $a'$ and integrating over $\theta$ in (\ref{eq:TotalInt}). The dependence of $W_{n,a,p_z=0,\zeta}^{h,\bar\sigma}$, summed over polarizations and helicities, on $n'$ and $a'$ is illustrated in \fig{fig:aprimeDependence}. There are two takeaways from this figure. First, the main contribution to the intensity comes from those $n'$ which are close to $n$. Second, for any particular $n'$, there is a relatively narrow region of $a'$ for which the intensity is significant. These features are general, although at larger $n$, smaller values of $n'$ become more relevant. Our numerical calculation searches for these narrow regions on the $a'$ axis for each $n' \leq n$, then sums only over those regions. This selection process causes the error in our calculation to grow with $n$, as there are more discarded terms, and for high enough $n$ the calculation of the Laguerre polynomials becomes numerically difficult. We also cut off our calculation when $a'$ approaches $N_{\rm caus}$, given by \eq{a39}. The results are independent of this cutoff (and the calculation rarely enforces this cutoff), as the intensity falls off quickly with increasing $a'$. Other features of \fig{fig:aprimeDependence} depend on specific parameters. For example, note that there are two peaks in intensity as a function of $a'$ for each $n'$ curve. In general, there are $a+1$ such peaks. Varying $qB$ and $\Omega$ changes the relative heights of the peaks.
 
\begin{figure}[t]
\begin{tabular}{cc}
      \includegraphics[width=3in]{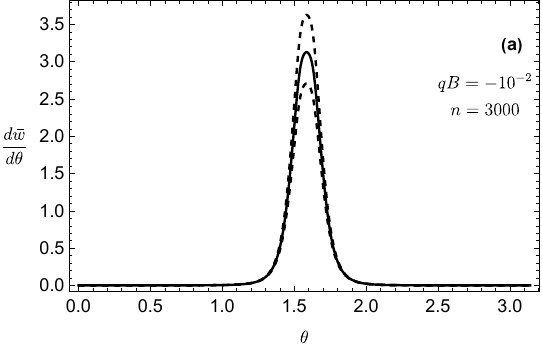}&
      \includegraphics[width=3in]{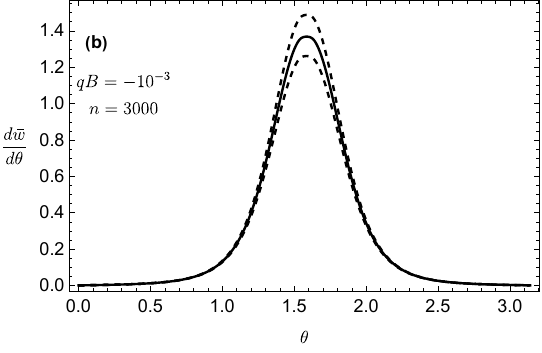} \\
      \includegraphics[width=3in]{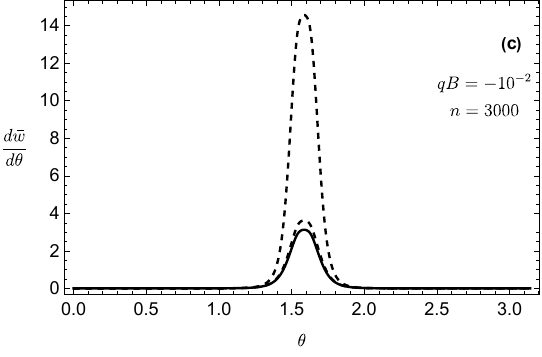}&
      \includegraphics[width=3in]{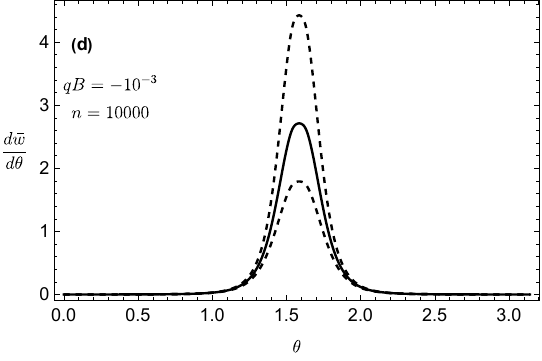} \\
\end{tabular}
  \caption{The intensity of photon emission $\overline{w} = \frac{W}{W_{\mathrm{cl}}}$ as a function of $\theta$ for indicated parameters. All 4 figures have $a = 1$. The solid curves: $\Omega = 0$. The upper (lower) dashed curves in (a), (b), and (d) $\Omega = 10^{-6}$ ($\Omega = - 10^{-6}$). In (c) the upper (lower) dashed curve has $\Omega = 10^{-5}$ ($\Omega = 10^{-6}$). Our units: $\hbar = c = M = 1$.
  }
\label{fig:ThetaIntensityComparison}
\end{figure}

The angular distribution of radiation peaks in the direction perpendicular to the magnetic field and vanishes along the rotation axis. These features, present in the classical and quantum non-rotating systems, are also salient in the rotating system as displayed in \fig{fig:ThetaIntensityComparison}. The novel feature is that depending on the sign of $\Omega$ the intensity is either enhanced ($\Omega>0$) or suppressed ($\Omega<0$) for the rotating fermion. This effect is enhanced at higher energies and for larger $|\Omega|$. As in the non-rotating case, most of the radiation by an ultrarelativistic fermion is concentrated in the cone, with small opening angle of the order $M/E\ll 1$ at large $E$.

It is instructive to compare our results to the radiation intensity of the non-rotating fermion. In the limit $\Omega\to 0$, the photon energy $\omega_0$ depends only on $n$ and $n'$, but not on $a$ and $a'$. This allows explicit summation over $a'$ in \eq{I1}, which can be performed  using the identity $\sum_{a'}I^2_{a,a'}(x)=1$ ~\cite{book:SokolovAndTernov}  and yields the well-known result for the synchrotron radiation intensity by a non-rotating fermion. If the fermion is ultrarelativistic $E/M\gg 1$ and the magnetic field is not very strong $|qB|/M^2\ll 1$ (implying in particular that $\omega_B\ll E$), then the spectrum becomes approximately continuous and one can employ the quasi-classical approximation which yields, for a non-rotating fermion \cite{LandauElectro},
\begin{align}\label{f1}
    W_\text{WKB}(\chi)= -\frac{q^2}{4\pi}\frac{\chi^2}{2}\int_0^\infty \frac{4+5\chi x^{3/2}+4\chi^2 x^3}{(1+\chi x^{3/2})^4}\mathrm{Ai}'(x)xdx\,,
\end{align}
where $\chi =|qB|E/M^3$ is a boost-invariant parameter and $\mathrm{Ai}$ is the Airy function. Quantum effects, such as fermion recoil, are negligible when $\chi\ll 1$, in which case (\ref{f1}) reduces to the classical expression $W_\mathrm{cl}$ given by \eq{d35}.
It is customary to present the radiation intensity in units of $W_\mathrm{cl}$.

\begin{figure}[t]
\begin{tabular}{cc}
      \includegraphics[height=3in]{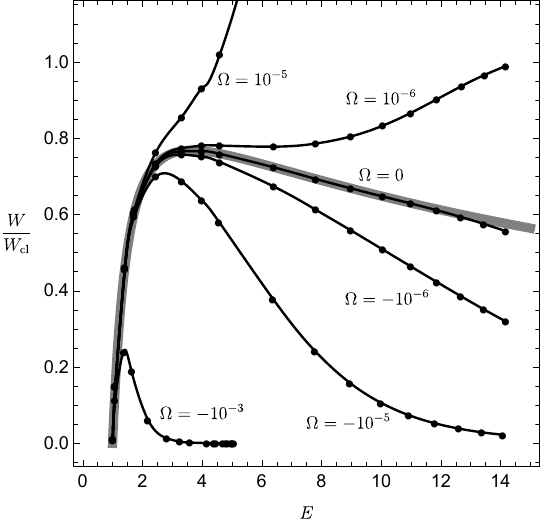} &
      \includegraphics[height=2in]{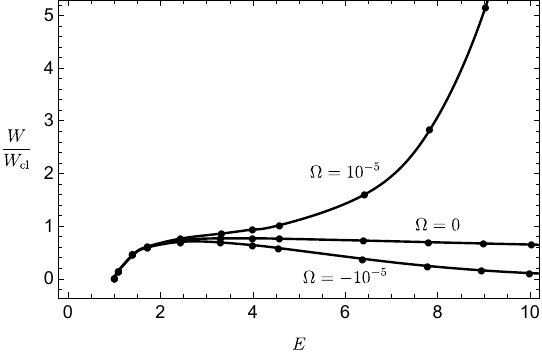}
\end{tabular}
  \caption{ The total intensity of the synchrotron radiation in units of the classical intensity \eq{d35} as a function of the initial energy $E$ at $qB=-0.01$. Left: Solid lines correspond to various angular velocities $\Omega$ and the thick gray line is the quasiclassical approximation at $\Omega=0$. The deviation of the result for $\Omega = 0$ from the quasiclassical result is due to loss of computational accuracy  for large $n$, as explained in the text. Right: An expanded view of the total intensity for $\Omega = 10^{-5}$. The dependence of the intensity on the initial value of $a$ is weak and not noticeable in the figure. Our units: $\hbar=c=M=1$.
  }
\label{fig:total1}
\end{figure}

\begin{figure}[t]
      \includegraphics[height=8cm]{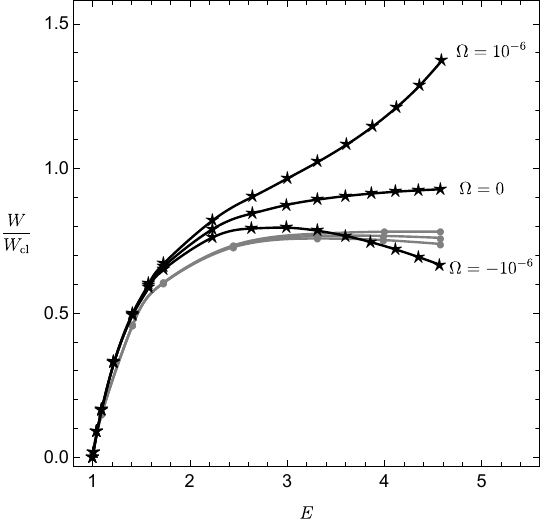}
  \caption{The same as in \fig{fig:total1} but with $\Omega = 0,\pm 10^{-6}$ and two values of $qB$. Grey lines with circles are for $qB = -10^{-2}$. Black lines with stars are for $qB = -10^{-3}$. Note the effect of rotation is greater at lower energies for the smaller field. Our units: $\hbar = c = M = 1$.}
\label{fig:total2}
\end{figure}

\fig{fig:total1} and \fig{fig:total2} show the intensity of the radiation for various values of the angular velocity  and the magnetic field. One can qualitatively understand the remarkably strong effect of slow rotation on intensity by considering an elementary classical model. A non-relativistic charged particle's trajectory is a  
combination of two circular motions: one with angular velocity $\Omega$ due to the rigid rotation, and another with angular velocity $\omega_B=|qB|/E$ due to the Lorentz force exerted by the magnetic field. The former is independent of the fermion energy $E$, whereas the latter decreases as $E^{-1}$. One can also notice that when the rotations due to the magnetic field and to the rigid rotation of the fermion are in the opposite direction (e.g.\ $qB<0$ and $\Omega<0$), the result is suppression of radiation. This happens because the rotating fermion experiences smaller effective $\omega_B$, hence smaller effective magnetic field \cite{Tuchin:2021lxl}. Conversely, when the two rotations are in the same direction (e.g.\ $qB<0$ and $\Omega>0$) we observe enhancement of the radiation. Finally, we notice a similarity between the classical  and quantum  models shown in \fig{figA3} and \fig{fig:total1} respectively.



\subsection{Dependence on \texorpdfstring{$N_\text{caus}$}{N caus}}
\begin{figure}[ht]
      \includegraphics[width=4.2in]{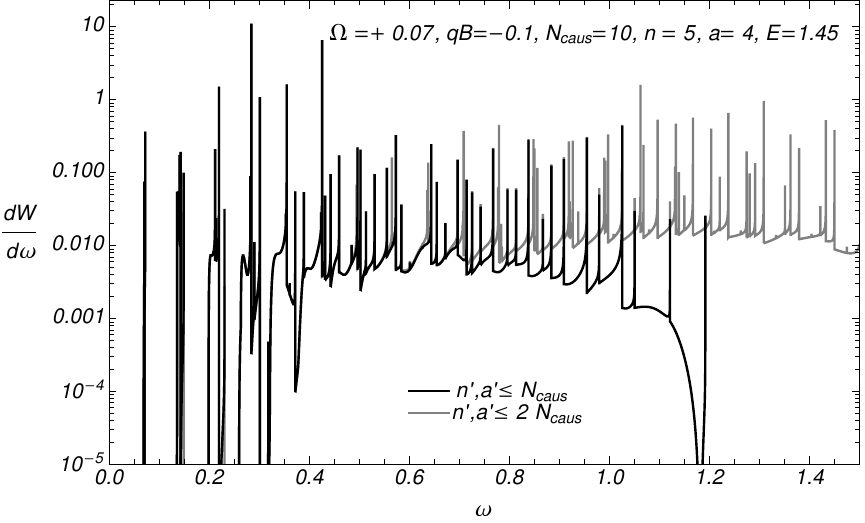}
  \caption{The spectrum of synchrotron radiation (\ref{eq:DiffIntensityInt}). Magnetic field strength is  $qB=-0.1$. Initial quantum numbers: $n=5$, $a=4$, $m=1/2$ and $p_z=0$, corresponding to $E=1.450$. The spectrum is summed over the final quantum numbers $a'$ and $n'$ up to $N_\text{caus}=10$ for the black lines and up to $2N_\text{caus}=20$ for the gray lines. Our units: $\hbar = c = M = 1$.}
\label{fig:spectrum_Ncaus}
\end{figure}

We checked that our results are not sensitive to moderate variation of the cutoff $N_\text{caus}$ given by \eq{a39}. The mathematical significance of the cutoff is to prevent fermion transitions from the initial state $E$ to the final state with large negative energy $E'$. This would occur at very large negative values of $m$, as seen implied by  \eq{dr1}. This divergence is a mere artifact of our neglect of the boundary condition at $r=1/\Omega$. Taking these conditions into account eliminates the divergence \cite{Buzzegoli:2023eeo}. The boundary conditions become important in rapidly rotating systems corresponding to region I in \fig{fig:qgp}. To illustrate this point we computed the spectrum of synchrotron radiation (\ref{eq:DiffIntensityInt}) by rapidly rotating fermion ($qB=-0.1 M^2$ and $\Omega = 0.07 M$). Fig.~\ref{fig:spectrum_Ncaus} shows significant dependence of the spectrum on the cutoff $N_\text{caus}$. Moreover, one can see that some photons are emitted with energies larger than the initial energy of the fermion. The emergence of these nonphysical states clearly indicates the breakdown of the `slow rotation' approximation and requires a careful treatment of the boundary conditions. This will be addressed in a dedicated paper.  

\section{Summary}\label{sec:summary}

In this paper we performed a detailed analysis of the synchrotron radiation by a fermion embedded in a uniformly rotating medium. Using the exact solution to the Dirac equation we analytically computed the radiation intensity spectrum given by \eq{eq:DiffIntensity},\eq{eq:theta0} and 
the total intensity given by Eqs~\eq{I1},\eq{eq:PhotonEnergyRotation},\eq{d22},\eq{d30},\eq{d31}. We also used these equations to study in \sec{sec:rad.posit.qB} the dependence of our results on the sign of the electric charge, direction of the magnetic field and the angular velocity. The final analytical results still contain a summation and an integration that we performed numerically.

Our main observation, exhibited in \fig{fig:total1}, is that rotation has a very strong effect on the radiation. This happens because the synchrotron frequency is inversely proportional to the fermion energy and therefore becomes comparable to the angular velocity of rotation at high energies, see Table~\ref{table} and \fig{fig:qgp}.  This is our benchmark result. It is completely general and can be applied to any area of physics where rotation is an important factor.

Of special interest are relativistic heavy-ion collisions (HIC), in which we can estimate the radiation produced by a single quark in the quark gluon plasma (QGP). Measurements of spin polarization provide a late-time vorticity of about $\Omega \sim 10^{-2} T \simeq 1$ MeV \cite{STAR:2017ckg} and estimates of the magnetic field give $|qB| \sim m^{2}_{\pi} = 2\times10^4$ MeV$^2$. Taking the temperature of the QGP to be $T = 300$ MeV and an effective thermal mass for the quarks of $M \sim T = 300$ MeV and adopting the mass units ($M=1$) used in this work, we have $(E,\, |qB|,\, \Omega) = (1\text{ to }3,\, 0.2,\, 0.003)$. We see that the energy and angular velocity of rotation in this scenario matches the values used in Fig. \ref{fig:total1} where $\Omega = 10^{-3}$. Instead, the magnetic field in the figure is lower ($|qB|= 10^{-2}$). However, the magnetic field intensity decreases to this magnitude at later stages of the QGP, and therefore the plot in Fig. \ref{fig:total1} is reasonably indicative of the expected radiation intensity of a quark embedded in the QGP
formed in HICs. Hence HICs are instances where the scales $\sqrt{|qB|},\, \Omega,$ and $\omega_{B}$ have similar orders of magnitude and thus synchrotron radiation is significantly enhanced or suppressed by rotation depending on the charge of the quark.

\acknowledgments
This work is partially supported by the US Department of Energy under Grants No.\ DE-FG02-87ER40371 and No.\ DE-SC0023692.

\appendix

\section{Polarization states}\label{app:polarization}

We would like to find the conserved quantities related to spin. To this end, we can construct the polarization operators in the following way \cite{book:SokolovAndTernov}. 
Let $\alpha$ be a $4\times 4$ matrix in the spinor space, e.g.\
$\alpha=\gamma^5,\vec{\Sigma},\sigma^{\mu\nu}F_{\mu\nu}$.
Given $\alpha$, we define $\tilde{\alpha}$ as
\begin{equation}
\tilde{\alpha}\equiv \{ H ,\, \alpha \} = H \alpha + \alpha H,
\end{equation}
where $H$ is the Hamiltonian of our system. With this definition it is easy to realize that
\begin{equation*}
    \left[ H, \tilde{\alpha} \right] = \left[ H^2, \alpha \right].
\end{equation*}
In this way, we reduce the problem of finding a quantity that commutes with the Hamiltonian
to finding a combination of gamma matrices that commutes with the Hamiltonian squared.
This is particularly easy for the free Dirac field (without EM field and rotation)
because
\begin{equation}
    H_0^2 = \b p^2 + M^2
\end{equation}
does not contain gamma matrices and every gamma matrix commutes with it.
A notable example is obtained if we choose $\alpha=\gamma^5$. In this case,
we can define
\begin{equation}
    h = \frac{1}{2M}\tilde{\gamma^5} = \frac{\vec\Sigma\cdot \vec{p}}{M}.
\end{equation}
This is the helicity operator.

\subsection{Helicity operator in external EM field}
In the presence of electromagnetic field, in order to maintain gauge invariance we
need to perform the minimal substitution:
\begin{equation*}
    H \to H - q\phi,\quad p_\mu \to P_\mu = p_\mu - q A_\mu.
\end{equation*}
Because of the gauge invariance we need to change the definition of $\tilde{\alpha}$
\begin{equation}
\tilde{\alpha}\equiv \{ H - q\phi ,\, \alpha \} .
\end{equation}
Using the fact that the Hamiltonian with electromagnetic field is
\begin{equation}
\label{eq:HEM}
H_{EM} = \gamma^0 \vec{\gamma}\cdot \vec{P} + \gamma^0 M + q\phi
    = \gamma^5 \vec{\Sigma}\cdot \vec{P} + \gamma^0 M + q\phi,
\end{equation}
%
choosing $\alpha=\gamma^5$, we obtain the helicity operator
\begin{equation}
h_{EM} = \frac{1}{2M}\tilde{\gamma^5}
    = \frac{1}{2M}\left[ (H_{EM} - q\phi ) \gamma^5 +  \gamma^5 (H_{EM} - q\phi ) \right]
    =\frac{\vec{\Sigma}\cdot \vec{P}}{M} .
\end{equation}
From (\ref{eq:HEM}), we also have
\begin{equation}
\vec{\Sigma}\cdot \vec{P} = \gamma^5 H_{EM} - \gamma^5 \gamma_0 M - q\phi \gamma^5,
\end{equation}
and we can write the helicity operator as
\begin{equation}
\label{eq:HelicityEMwithH}
h_{EM} =\frac{ \gamma^5 H_{EM} - \gamma^5 \gamma^0 M - q\phi \gamma^5}{M}.
\end{equation}
From the equation above we find by a straightforward  calculation that
\begin{equation}
\label{eq:hemTimederivative}
\frac{d}{dt} h_{EM} = \frac{\de}{\de t} h_{EM} + \I \left[ H_{EM}, h_{EM}\right]
    = -\frac{q}{M}\left(\frac{\de\vec{A}}{\de t} + \vec{\nabla}\phi \right)\cdot\vec{\Sigma}
    = \frac{q}{M}\vec{\Sigma}\cdot \vec{E}.
\end{equation}
We see then that the helicity operator is a constant of motion (and commutes with the
Hamiltonian) if $\vec{E}=0$.

\subsection{Helicity operator in external electromagnetic field and in rotating frame}

In the rotating frame we can perform the same operation to obtain the spin operators. This time we have to define the quantity
\begin{equation}
\tilde{\alpha}\equiv \{ H_\Omega - q\phi - \vec{\Omega}\cdot\vec{J},\, \alpha \} ,
\end{equation}
where denoted the Hamiltonian as $H_\Omega$ for the sake of notation consistency in this Appendix. In the main part of the paper it is denoted simply as $H$. Explicitly,
\begin{equation}
\label{eq:HOmega}
H_\Omega = \gamma^0 \vec{\gamma}\cdot \vec{P} + \gamma^0 M + q\phi + \vec{\Omega}\cdot\vec{J}
    = \gamma^5 \vec{\Sigma}\cdot \vec{P} + \gamma^0 M + q\phi + \vec{\Omega}\cdot\vec{J}.
\end{equation}

Considering the helicity operator  $\tilde{\gamma^5}$, we find that it has the same form as in the external electromagnetic field:
\begin{equation*}
\frac{\tilde{\gamma^5}}{2M}
    = \frac{1}{2M}\left[ (H_\Omega - q\phi - \vec{\Omega}\cdot\vec{J}) \gamma^5
    +  \gamma^5 (H_\Omega - q\phi - \vec{\Omega}\cdot\vec{J}) \right]
    = \frac{\vec{\Sigma}\cdot\vec{P}}{M}
    = h_{EM} .
\end{equation*}
In terms of the Hamiltonian (\ref{eq:HOmega}) the helicity operator can be written as
\begin{equation}
\label{eq:HelicityEMwithOmega}
M h_{EM} = \gamma^5 \left(H_\Omega - \vec{\Omega}\cdot\vec{J}\right) -\gamma^5 \gamma^0\, M
    - \gamma^5  q\phi.
\end{equation}
Even in this case we have:
\begin{equation}
\label{eq:hemTimederivativeOmega}
\frac{d}{dt} h_{EM} = \frac{\de}{\de t} h_{EM} + \I \left[ H_\Omega, h_{EM}\right]
    = -\frac{q}{M}\left(\frac{\de\vec{A}}{\de t} + \vec{\nabla}\phi \right)\cdot\vec{\Sigma}
    = \frac{q}{M}\vec{\Sigma}\cdot \vec{E}.
\end{equation}
Thus in a purely magnetic background, helicity is conserved. 
We can also check that
\begin{equation}
\label{eq:Commhem_OmegaJ}
\left[ h_{EM},\, \vec{\Omega}\cdot\vec{J}\right] = 0.
\end{equation}

\subsection{Canonical spin tensor - spin magnetic and electric moments}
The canonical spin tensor is
\begin{equation}
S^{\lambda,\mu\nu} = \bar\psi
    \frac{\I}{8}\left\{ \gamma^\lambda,\,\left[ \gamma^\mu,\,\gamma^\nu\right]\right\}\psi .
\end{equation}
For $\lambda=0$, we consider the gamma matrices:
\begin{equation}
\alpha^{ij}\equiv \frac{\I}{4}\left\{ \gamma^0,\,\left[ \gamma^i,\,\gamma^j \right]\right\} =
    \frac{\I}{2}\gamma^0 \left[ \gamma^i,\,\gamma^j \right] = \gamma^0 \sigma^{ij}.
\end{equation}
Starting from this we define the spin magnetic vector $\vec{\mu}$ as
\begin{equation}
\label{eq:defspinmagnetic}
\mu_i = \epsilon_{ijk} \tilde{\alpha}^{jk}
    = \frac{\epsilon_{ijk}}{2M}\left[ (H_\Omega - q\phi - \vec{\Omega}\cdot\vec{J}) \alpha^{jk}
    +  \alpha^{jk} (H_\Omega - q\phi - \vec{\Omega}\cdot\vec{J}) \right].
\end{equation}
We find that \cite{Schwinger:1948iu,Sommerfield:1957zz}
\begin{equation}
\vec{\mu}= \vec{\Sigma} - \frac{\I\gamma^0 \gamma^5}{2M}\vec{\Sigma}\times (\vec{p}-q\vec{A})
\end{equation}
and that
\begin{equation}
\frac{d}{d t}\vec{\mu}= \gamma^0\vec{\Sigma}\times q\vec{B} - \I\gamma^0 \gamma^5\vec{\Sigma}\times q\vec{E}.
\end{equation}
In particular, for $\vec{E}=0$ and $\vec{B}= B\, \vec{\hat{z}}$ we see that $\mu_z$ is a constant of
motion. In this case using
\begin{equation*}
\alpha^{12} = \gamma^0 \Sigma^z,
\end{equation*}
and
\begin{equation}
\gamma^0 \sigma^z \left(H_\Omega - \vec{\Omega}\cdot\vec{J}\right)= 
    \left(H_\Omega - \vec{\Omega}\cdot\vec{J}\right)\gamma^0 \Sigma^z - 2\, \, P_z \gamma^0 \gamma^5 ,
\end{equation}
we can write:
\begin{equation}
\begin{split}
\mu_z = & \frac{1}{2M}\left[ (H_\Omega - \vec{\Omega}\cdot\vec{J}) \gamma^0 \Sigma^z
    +  \gamma^0 \Sigma^z (H_\Omega - \vec{\Omega}\cdot\vec{J}) \right]
= \frac{1}{M}\left[(H_\Omega - \vec{\Omega}\cdot\vec{J})\gamma^0\Sigma^z
    - P_z \gamma^0 \gamma^5\right]\\
=& \frac{1}{M}\left[\left( H_\Omega - \vec{\Omega}\cdot\vec{J}\right) \I \gamma^0 \gamma^1\gamma^2
    - P_z \gamma^0 \gamma^5\right).
\end{split}
\end{equation}

\bibliographystyle{apsrev4-1}
\bibliography{biblio}
\end{document}